%% file: main.tex
\documentclass[runningheads]{llncs}

\usepackage{graphicx}
\usepackage{enumerate}
\usepackage{amsmath}
\usepackage{amssymb}
\usepackage{tikz}
\usepackage{caption}
\usepackage{subcaption}
\usepackage{algpseudocode}
\usepackage{algorithm}
\usepackage{mathtools}
\usepackage{hyperref}
\usepackage{pifont}

\usetikzlibrary {graphs}
\usetikzlibrary{positioning}
\usetikzlibrary {arrows.meta}
\usetikzlibrary{shapes.geometric}
\usetikzlibrary{backgrounds}

\newcommand{\tlaplus}{TLA\textsuperscript{+}}

\definecolor{darkishgreen}{RGB}{0,150,0} 

\newcommand{\xmark}{\ding{55}}%

\newcommand*{\true}{\text{True}}
\newcommand*{\false}{\text{False}}

\AtBeginDocument{%
  \providecommand\BibTeX{{%
    \normalfont B\kern-0.5em{\scshape i\kern-0.25em b}\kern-0.8em\TeX}}}

\begin{document}

\tikzset{>=latex}

\newcommand{\todo}[1]{\textcolor{red}{TODO: #1}}


\newcommand{\pardesc}[1]{}



\title{Interactive Safety Verification of Distributed Protocols by Inductive Proof Decomposition}

\author{William Schultz\inst{1} \and
Edward Ashton\inst{2} \and Heidi Howard\inst{2} \and
Stavros Tripakis\inst{1}}

\institute{Northeastern University, Boston, USA \and
Microsoft Research, Cambridge, UK
}

\titlerunning{Verification of Distributed Protocols by Inductive Proof Decomposition}





\maketitle

\begin{abstract}
    Many techniques for the automated verification of distributed protocols have been developed over the past several years, but their performance is still unpredictable and their failure modes can be opaque for industrial scale verification tasks. 
    Thus, in practice, large-scale verification efforts typically require some amount of human guidance. 
    In this paper, we present \textit{inductive proof decomposition}, a new methodology for interactive safety verification that provides a compositional, interactive approach to inductive invariant development. Our approach guides the human-aided development of inductive invariants via a novel structure, an \textit{inductive proof graph}, which is built incrementally by a human verifier, working backwards from a target safety property. 
    A user is guided by induction counterexamples that are localized to specific nodes of this graph, and nodes of this proof graph are further decomposed based on logical actions that appear in a protocol's transition relation.   
    Our decomposition also enables a localized \textit{variable slicing} technique that hides irrelevant protocol state at each sub-component of an inductive proof, allowing a user to focus on fine-grained sub-problems rather than a large, monolithic inductive invariant.
    We present our technique and experience applying it to develop inductive safety proofs of several complex distributed protocols, including the Raft consensus protocol, which is beyond the capabilities of modern automated verification tools. We also demonstrate how the developed proof graphs provide additional insight into the structure of a protocol proof and its correctness.
\end{abstract}

\section{Introduction}

\pardesc{Automated verification tools have continued to scale, but still a jagged boundary at their limits, esp. for industrial scale verification problems}

Verifying the safety of large-scale distributed and concurrent systems remains an important and difficult challenge. 
These protocols serve as the foundation of many modern fault-tolerant systems, making the correctness of these protocols critical to the reliability of large scale database and cloud systems \cite{2020cockroachdb,2020tidb,2012spannerosdi}. 
Formally verifying the safety of these protocols typically centers around development of an \textit{inductive invariant}, an assertion about system state that is preserved by all protocol transitions. 
Developing inductive invariants, however, is one of the most challenging aspects of safety verification and has typically required a large amount of human effort for real world protocols \cite{wilcox2015verdi,woos2016planning}.

Over the past several years, particularly in the domain of distributed protocol verification, there have been several recent efforts to develop more automated inductive invariant development techniques \cite{2021ic3posymmetry,yao2021distai,2016padonivy,koenigpadon2022}. 
Many of these tools are based on modern model checking algorithms like IC3/PDR \cite{2020firstorderquantified,koenigpadon2022,2021ic3posymmetry,Goel2021TowardsAA,2017updr}, and others based on syntax-guided or enumerative invariant synthesis methods \cite{2021swisshance,YaoTGN22}. 
These techniques have made significant progress on solving various classes of distributed protocols, including some variants of real world protocols like the Paxos consensus protocol \cite{lamport2001paxos,Goel2021TowardsAA}.
The theoretical complexity limits facing these techniques, however, limit their ability to be fully general \cite{2016padondecidability} and, even in practice, the performance of these tools on complex protocols is still unpredictable, and their failure modes can be opaque.
In particular, a key drawback of these methods is that, in their current form, they are very much ``all or nothing''. 
That is, if they solve a given problem, no manual proof effort is needed, but if a problem falls outside the scope of what they can solve, little assistance is provided in terms of how to develop a manual proof or how a human can offer guidance to the tool. 

In practice, real world, large-scale verification efforts often require some amount of human interaction i.e., a human provides guidance when an automated engine is unable to automatically prove certain properties about a design or protocol. 
For example, recent verification efforts of industrial scale protocols either note the high amount of human effort in developing inductive invariants or leave them as future goals \cite{schultz2021formal,braithwaite2020formal}. 
Several recent, automated approaches have also adopted a paradigm of integrating human assistance to accelerate proofs for larger verification problems e.g., in the form of a manually developed refinement hierarchy \cite{Goel2021TowardsAA,Ma2022SiftUR}.

\pardesc{There is lack of prior focus on techniques for human-aided proof development when tools fail.}




\pardesc{Prior attempts at addressing this problem via interactive approaches}
Though there has been a large amount of work on scaling \textit{automated} protocol verification techniques, there has been considerably less focus on \textit{interactive} verification. That is, consideration of how a human can proceed effectively with an inductive proof when a tool fails to solve a verification task automatically. 
The Ivy framework \cite{2016padonivy} was a notable, more recent attempt to address the interactive safety verification problem, but its techniques were targeted at specific goals which only addressed partial aspects of the problem. 
Namely, their focus was primarily on (1) ensuring decidability of verification conditions and (2) incorporating a human into the loop of counterexample generalization heuristics. 
Though this addressed some aspects of the human-machine verification interface, it did not consider other, key issues that arise in large-scale inductive proof efforts. For example, it did not consider how to manage the structure of a large inductive invariant effectively as it is being developed, how to provide feedback to a user about progress on the proof, or how to effectively allow localized reasoning on sub-components of a larger proof, etc. 

In addition to frameworks like Ivy, there is a large amount of work on the use of interactive theorem proving for system verification \cite{2015ironfleet,2017choikami} e.g., using systems like Coq \cite{bertot2013interactive}, Isabelle/HOL \cite{nipkow2002isabelle}, ACL2 \cite{hunt2017industrial}, etc. The learning curve for these tools is typically steep, though, and they have typically offered a significantly lower degree of automation, making them more laborious to use for many verification efforts and for protocol designers or engineers \cite{newcombe2014amazon}. Thus, although these tools provide a relatively high degree of interactivity, they are often quite complex and tedious to use for practical verification efforts. 



\pardesc{In this paper we propose a new, interactive safety verification methodology, ``inductive proof decomposition"}
In this paper we present a new, interactive safety verification methodology, \textit{inductive proof decomposition}, that provides a compositional approach to interactive inductive invariant development, enabling a smooth integration between human effort and machine guidance for large-scale safety proof efforts. 
We are focused on the human-aided development of inductive invariants, typically with the assistance of a backend solver for checking inductive proof obligations that can provide counterexamples to induction. Standard approaches to this process (e.g., as in the paradigm of Ivy) essentially proceed by having a human verifier examine induction counterexamples in a linear fashion, with a goal of constructing a monolithic list of lemma invariants whose conjunction form a valid inductive invariant. We argue that this standard model is poorly suited for large and complex verification efforts, where inductive invariants may grow to include potentially dozens of complex conjuncts about protocol state and individual counterexamples become increasingly complex to analyze. Our technique aims to make this process fundamentally \textit{compositional}, which we believe is a core aspect of any ``intuitive" proof process undertaken by a human (e.g. as in ``pen and paper" style proofs), and is also crucial for managing complexity in any large proof effort \cite{lamport1995write,WilcoxThesis21,taube2018modularity}.

Our technique is based around a novel, formal structure which we introduce and define, the \textit{inductive proof graph}, which imposes a particular compositional structure on an inductive invariant, while also taking into account the distinct logical actions that are present in most concurrent and distributed protocols. 
This graph structure makes explicit the relative induction dependencies between lemmas of a monolithic inductive invariant, and our proof methodology guides a human verifier to incrementally construct this graph structure by working backwards from a target safety property. 
Throughout the process, machine guidance is provided in the form of relative induction counterexamples that are maintained at each node of this graph, and so can be reasoned about locally, rather than considered in the scope of the entire inductive invariant. 
This also allows for local abstractions to be applied that reduce the complexity of counterexamples to be examined. 
In particular, our approach includes a novel \textit{variable slicing} technique, that projects out protocol variables that are irrelevant to a local node of the proof graph, often significantly reducing the scope of information presented to a user, easing their analysis task.


\pardesc{Evaluation of our technique}
We apply our technique to several distributed protocols, including a large, industrial-scale specification of the Raft \cite{raftpaper} consensus protocol, demonstrating the effectiveness of our technique and its ability to allow human verifiers to work with large proof structures effectively. We also show how the resulting proof graph artifacts provide additional insight into protocol correctness.

In summary, our contributions are as follows:
\begin{itemize}

    \item Definition and formalization of \textit{inductive proof graphs}, a formal structure representing the logical dependencies between conjuncts of an inductive invariant and actions of a distributed or concurrent protocol. (Section \ref{sec:ind-proof-graphs})
    \item \textit{Inductive proof decomposition}, a methodology for large scale interactive safety proofs that is based around the incremental, counterexample-guided construction of an inductive proof graph. (Section \ref{sec:ind-proof-decomp})
    \item Implementation of our technique in an interactive verification tool, \mbox{\textsc{Scimitar}}, and evaluation of our method on several distributed protocols, including a large-scale specification of the Raft \cite{raftpaper} protocol. (Section \ref{sec:evaluation})
\end{itemize}

\newcommand{\ca}[1]{\textcolor{black}{#1}}
\newcommand{\cb}[1]{\textcolor{black}{#1}}
\newcommand{\cc}[1]{\textcolor{black}{#1}}
\newcommand{\cd}[1]{\textcolor{black}{#1}}

\section{Preliminaries}
\label{sec:background}

In this paper we are focused on the problem of safety verification of protocols formalized as discrete transition systems, which consists of a core problem of finding adequate inductive invariants. Furthermore, we are focused on verification of systems that are assumed to be correct i.e., we assume various bug-finding methods (\cite{holzmann1997model},\cite{biere03}) have been applied upfront before a proof is undertaken.


\paragraph{Transition Systems and Invariants}
\label{sec:prelim-transition-systems}

The protocols considered in this paper are modeled as {\em symbolic transition systems}, where a state predicate {\em I} defines the possible values of state variables at initial states of the system, and a predicate $T$ defines the {\em transition relation}. A transition system $M$ is then defined as $M=(I, T)$, and the \textit{behaviors} of $M$ are defined as the set of all sequences of states $\sigma_1 \rightarrow \sigma_2 \rightarrow \dots$ that begin in some state satisfying $I$ and where every transition $\sigma_i \rightarrow \sigma_{i+1}$ satisfies $T$. The \textit{reachable states} of $M$ are the set of all states that exist in some behavior. In this paper we are concerned with the verification of \textit{invariants}, which are predicates over the state variables of a system that hold true at every reachable state of a system $M$. In this paper, we also assume that the transition relation $T$ for a system $M$ is composed of distinct logical actions, $T = A_1 \vee \dots \vee A_k$. For example, a simple transition relation of this form is $T = (x' = x+ 1) \vee (x' = x + 2)$, where a primed state variable ($x'$) represents the value of that state variable in the next state. 


We also define a restricted class of transition systems where transition relations are expressed in a \textit{guarded action} style. That is, systems where all actions $A$ are of the form $A = Pre \wedge Post$, where $Pre$ is a predicate over current state variables and $Post$ is a conjunction of update formulas of the form $x_i' = f_i(\mathcal{D}_i)$, where $f_i$ is some expression over a subset of current state variables $\mathcal{D}_i$. For simplicity, we assume that all state variables always appear in $Post$, and that for variables unchanged by a protocol action, they simply appear in $Post$ with an identity update expression, $x_i' = x_i$. Note that although systems in guarded action style have deterministic update expressions, these systems can still be non-deterministic, due to non-determinism over constant system parameters, e.g., as illustrated in Figure \ref{fig:simple-consensus-spec}, which describes a simple, leader-based distributed consensus protocol.





\begin{figure}[t]
    \label{fig:simple-consensus-inv-def}
    \tiny
    \fontsize{8}{9.5}\selectfont
    \hrule
    \begin{subfigure}[b]{.523\textwidth}
    \setlength{\fboxrule}{0pt}
    \fbox{
    \begin{minipage}[t]{0.45\textwidth}
    \begingroup
    \addtolength{\jot}{-0.48em}
    \vspace{-0.25cm}
    \begin{align*}
    &\text{CONSTANTS } \,\, Node, Value, Quorum \\[0.6em]
    &\text{VARIABLES} \,\,  voteReqMsg, voted, \\ &voteMsg, votes, leader, decided\\[0.7em]
        &\text{\textcolor{gray}{Protocol actions.}}\\[0.3em]
        &\mathit{SendRequestVote}(src, dst) \triangleq\\
            & \hspace{3pt} \wedge voteReqMsg' = voteReqMsg \cup \{\langle src, dst \rangle\} \\[0.5em]
        &\mathit{SendVote}(src, dst) \triangleq \\
            & \hspace{3pt} \wedge \neg voted[src] \\
            & \hspace{3pt} \wedge \langle dst,src \rangle \in voteReqMsg \\
            & \hspace{3pt} \wedge voteMsg' = voteMsg \cup \{\langle src,dst \rangle\} \\
            & \hspace{3pt} \wedge voted'[src] := \true \\ 
            & \hspace{3pt} \wedge voteReqMsg' = voteReqMsg \setminus \{\langle src,dst \rangle\} \\[0.5em]
        &\mathit{RecvVote}(n, sender) \triangleq \\
            & \hspace{3pt} \wedge \langle sender,n \rangle \in \mathit{voteMsg} \\
            & \hspace{3pt} \wedge \mathit{votes'}[n] := \mathit{votes}[n] \cup \{sender\} \\[0.5em]
        &\mathit{BecomeLeader}(n, Q) \triangleq \\
            & \hspace{3pt} \wedge Q \subseteq \mathit{votes}[n] \\
            & \hspace{3pt} \wedge \mathit{leader'}[n] := \true \\[0.5em]
        &\mathit{Decide}(n, v) \triangleq \\
            & \hspace{3pt} \wedge \mathit{leader}[n] \\
            & \hspace{3pt} \wedge \mathit{decided}[n] = \{\} \\
            & \hspace{3pt} \wedge \mathit{decided'}[n] := \{v\}
    \end{align*}
    \endgroup
    \end{minipage}
    }
    \end{subfigure}
    \begin{subfigure}[b]{.3\textwidth}
    \setlength{\fboxrule}{0pt}
    \fbox{
    \begin{minipage}[t]{0.95\textwidth}
        \begingroup
        \addtolength{\jot}{-0.4em}
        \vspace{-0.25cm}
        \begin{align*}
            &\mathit{NoConflictingValues} \triangleq 
            \text{\fcolorbox{green!60!black}{green!10!white}{\scriptsize\makebox[0.84in][c]{\textcolor{green!50!black}{Safety property}}}} \\
            &\hspace{10pt} \forall n_1,n_2 \in Node, v_1,v_2 \in Value : \\
                & \hspace{18pt} (v_1 \in decided[n_1] \wedge v_2 \in decided[n_2]) 
                 \\& \,\,\,\,\,\quad\quad  \Rightarrow (v_1 = v_2) \\[-0.7em]\\
            \\
            &\mathit{Uni}\mathit{queLeaders} \triangleq \\
            &\,\,\,\,\forall n_1,n_2 \in Node : \\
            &\,\,\,\,\,leader[n1] \wedge leader[n_2] \Rightarrow (n_1 = n_2)\\[0.4em]
            &\mathit{Leader}\mathit{HasQuorum} \triangleq \\
            &\,\,\,\, \forall n \in Node : leader[n] \Rightarrow \\ \,\,
            &\,\,\,\, (\exists Q \in Quorum : votes[n] = Q)
            \\[0.4em]
            &\mathit{LeadersDecide}\triangleq\\ &\,\,\,\,\forall n \in Node :  \\ & \,\,\,\,(decided[n] \neq \{\}) \Rightarrow leader[n]
        \end{align*}
        \vspace{-0.73cm}
        \begin{align*}
            %
            %
            \mathit{Ind} &\triangleq 
               \text{\fcolorbox{gray!80}{gray!15}{\scriptsize\rlap{\phantom{\makebox[1.29in]{\ }}}
               \parbox[c][0.73em][c]{1.15in}{\centering\textcolor{gray!60!black}{Inductive invariant.}}}}\\
            &\wedge \mathit{NoConflictingValues} \\
            &\wedge \mathit{UniqueLeaders} \\
            &\wedge \mathit{LeaderHasQuorum} \\
            &\wedge \mathit{LeadersDecide} \\
            & \wedge \mathit{NodesVoteOnce} \\
            & \wedge \mathit{VoteRecvdImpliesVoteMsg} \\
            & \wedge \mathit{VoteMsgsUnique} \\
            & \wedge \mathit{VoteMsgImpliesVoted}
        \end{align*}
        \endgroup
    \end{minipage}
    }
    \end{subfigure}
    \vspace{0.2em}
    \hrule
    \caption{Summarized specification of \emph{SimpleConsensus}, a simple leader-based consensus protocol. The protocol is parameterized on a finite set of nodes ($Node$) and values ($Value$), and nodes elect one leader to decide on a value. The safety property ($NoConflictingValues$) is an invariant stating that no two nodes can decide conflicting values. The associated inductive invariant, $Ind$, is used to establish this safety property.
    }
    \label{fig:simple-consensus-spec}
\end{figure}

\paragraph{Inductive Invariants and Relative Induction}

The standard technique for proving an invariant $S$ of a system $M=(I,T)$ is to develop an {\em inductive invariant}~\cite{mannasafetybook}, which is a state predicate $Ind$ such that $Ind \Rightarrow S$ and
\begin{align}
    &I \Rightarrow Ind \label{cond-init} \\
    &Ind \wedge T \Rightarrow Ind' \label{cond-next}
\end{align}
where $Ind'$ denotes the predicate $Ind$ where state variables are replaced by their primed, next-state versions.
Conditions~(\ref{cond-init}) and~(\ref{cond-next}) are, respectively, referred to as \textit{initiation} and \textit{consecution}. 
Condition~(\ref{cond-init}) states that $Ind$ holds at all initial states.
Condition~(\ref{cond-next}) states that $Ind$ is {\em inductive}, i.e., if it holds at some state $s$ then it also holds at any successor of $s$.
Together these two conditions imply that $Ind$ is also an invariant, i.e., that $Ind$ holds at all reachable states.

Typically, an inductive invariant is represented as a strengthening of $S$ via a conjunction of smaller \textit{lemma invariants}, $L_1,\dots,L_k$, such that the final inductive invariant is defined as $Ind = S \wedge L_1 \wedge \dots \wedge L_k$. Throughout this paper we assume inductive invariants can be represented in this form. Note also that for a given system $M=(I,T)$, a state predicate may be inductive only under the assumption of some other predicate. For given state predicates $Ind$ and $L$, if the formula $L \wedge Ind \wedge T \Rightarrow Ind'$ is valid, we say that $Ind$ is \textit{inductive relative to $L$}. 




\section{Inductive Proof Graphs}
\label{sec:ind-proof-graphs}

Our inductive invariant development technique is based around a core logical data structure, the \textit{inductive proof graph}, which we discuss and formalize in this section. This graph encodes the structure of an inductive invariant in a way that is amenable to localized reasoning and human interpretability, and also to integration of machine assistance, as we discuss further in Sections \ref{sec:ind-proof-decomp} and \ref{sec:evaluation}.

\subsection{Decomposing Inductive Invariants}

A \textit{monolithic} approach to inductive invariant development, where one searches for a single inductive invariant that is a conjunction of smaller lemmas, is a general proof methodology for safety verification \cite{mannasafetybook}. Any monolithic inductive invariant, however, can alternatively be viewed in terms of a \textit{relative induction} dependency structure, which is the initial basis for our formalization of inductive proof graphs, and which decomposes an inductive invariant based on this structure.

Namely, for a transition system $M=(I,T)$ and associated invariant $S$, given an inductive invariant
\begin{align*}
    Ind = S \wedge  L_1 \wedge \dots \wedge L_k
\end{align*}
each lemma in this overall invariant may only depend inductively on some other subset of lemmas in $Ind$. More formally, proving the consecution step of such an invariant requires establishing validity of the following formula
\begin{align}
    (S \wedge L_1 \wedge \dots \wedge L_k) \wedge T \Rightarrow (S \wedge L_1 \wedge \dots \wedge L_k)'
\end{align}
which can be decomposed into the following set of independent proof obligations:
\begingroup
\addtolength{\jot}{-0.2em}
\begin{align}
    \begin{split}
        \label{eq:ind-formula-decomp}
        (S \wedge L_1 \wedge \dots \wedge L_k) &\wedge T \Rightarrow S' \\
        (S \wedge L_1 \wedge \dots \wedge L_k) &\wedge T \Rightarrow L_1' \\
        &\vdots \\[0.0em]
        (S \wedge L_1 \wedge \dots \wedge L_k) &\wedge T \Rightarrow L_k'
    \end{split}
\end{align}
\endgroup
If the overall invariant $Ind$ is inductive, then each of the proof obligations in Formula \ref{eq:ind-formula-decomp} must be valid. That is, we say that each lemma in $Ind$ is inductive \textit{relative} to the conjunction of lemmas in $\{ S, L_1, \dots, L_k\}$.

With this in mind, if we define $\mathcal{L} = \{ S, L_1, \dots, L_k\}$ as the lemma set of $Ind$, we can consider the notion of a \textit{support set} for a lemma in $\mathcal{L}$ as any subset $U \subseteq \mathcal{L}$ such that $L$ is inductive relative to the conjunction of lemmas in $U$ i.e.,
        $\left( \bigwedge_{\ell \in U} \ell \right) \wedge L \wedge T \Rightarrow L'$.
As shown above in Formula \ref{eq:ind-formula-decomp}, $\mathcal{L}$ is always a support set for any lemma in $ \mathcal{L}$, but it may not be the smallest support set. 
This support set notion gives rise to a structure we refer to as the \textit{lemma support graph}, which is induced by each lemma's mapping to a given support set, each of which may be much smaller than $\mathcal{L}$.

For distributed and concurrent protocols, the transition relation of a system $M=(I,T)$ is most typically a disjunction of several distinct actions i.e., $T=A_1 \vee \dots \vee A_n$, as shown in the example of Figure \ref{fig:simple-consensus-spec}.
So, each node of a lemma support graph can be augmented with sub-nodes, one for each action of the overall transition relation. Lemma support edges in the graph then run from a lemma to a specific action node, rather than directly to a target lemma. 
Incorporation of this action-based decomposition now lets us define the full inductive proof graph structure.
%
%
\begin{definition}
    For a system $M=(I,T)$ with $T=A_1 \vee \dots \vee A_n$, an \textit{inductive proof graph} is a directed graph $(V,E)$ where 
    \begin{itemize}
        \item 
        $V = V_{L} \cup V_{A}$ consists of a set of \textit{lemma nodes} $V_L$ and \textit{action nodes} $V_A$, where 
        \begin{itemize}
            \item $V_L$ is a set of state predicates over $M$.
            \item $V_A = V_L \times \{A_1,\dots,A_n\}$ is a set of action nodes, associated with each lemma node in $V_L$.
        \end{itemize}
        
        \item $E \subseteq V_L \times V_A$ is a set of \textit{lemma support edges}.
    \end{itemize}
\end{definition} 


Figure \ref{fig:abs-ind-proof-with-actions} shows an abstract example of an inductive proof graph along with its corresponding inductive proof obligations annotating each action node. For simplicity, when depicting inductive proof graphs, if an action node is self-inductive, we omit it. Similarly, action nodes are always associated with a particular lemma, so we visualize edges connecting action nodes to their parent lemma node, even though these are not edges in the formal definition. 

\tikzset{%
    abstractlemmaedge/.style={},
    abstractactionedge/.style={}
}

\tikzset{%
    lemmanode_abstract/.style={rounded corners, draw=black, thick, minimum size=7.5mm,font=\large},
    actionnode_abstract/.style={fill=gray!20},
}

\begin{figure}[t]
    \small
        \centering
        \newcommand{\minwidth}{9.1mm}
        \scalebox{0.76}{
        \begin{tikzpicture}
            \node (S) [draw,minimum width=\minwidth,lemmanode_abstract] {$S$};
            \node[above left=0.5cm of S] (T1) [draw,actionnode_abstract] {\small$A_1$};
            \node[above right=0.5cm of S] (T2) [draw,actionnode_abstract] {\small$A_2$};
            \node[above left=0.1cm and 0.25cm of T1] (L1) [draw,minimum width=\minwidth,lemmanode_abstract] {$L_{1}$};
            \node[above right=1.25cm and 1.125cm of L1] (L12) [draw,lemmanode_abstract,minimum width=\minwidth] {$L_{1.2}$};
            \node[above right=0.1cm and 0.25cm of T2] (L2) [draw,lemmanode_abstract,minimum width=\minwidth] {$L_{2}$};
            \node[above=0.4cm of L1] (L1_T1) [draw,actionnode_abstract]{\small$A_{1}$};
            \node[above=0.4cm of L2] (L2_T2) [draw,actionnode_abstract]{\small$A_{1}$};
            \node[above left=0.2cm and 0.8cm of L1_T1] (L11) [draw,lemmanode_abstract,minimum width=\minwidth] {$L_{1.1}$};

            \node[left=0.1cm of T1] (O1) {\footnotesize $(L_1 \wedge S \wedge A_1 \Rightarrow S')$};
            \node[right=0.1cm of T2] (O2) {\footnotesize $(L_2 \wedge S \wedge A_2 \Rightarrow S')$};
            \node[left=0.1cm of L1_T1] (O3) {\footnotesize $(L_{1.1} \wedge L_{1.2} \wedge L_1 \wedge A_1 \Rightarrow L_1')$};
            \node[right=0.1cm of L2_T2] (O4) {\footnotesize $(L_{1.2} \wedge L_2 \wedge A_1 \Rightarrow L_2')$};

            \draw[->,abstractactionedge,densely dotted] (L1_T1) -- (L1);
            \draw[->,abstractactionedge,densely dotted] (L2_T2) -- (L2);
            \draw[->,abstractlemmaedge] (L11) -- (L1_T1);
            \draw[->,abstractlemmaedge] (L12) -- (L1_T1);
            \draw[->,abstractlemmaedge] (L12) -- (L2_T2);
            \draw[->,abstractlemmaedge] (L1) -- (T1);
            \draw[->,abstractlemmaedge] (L2) -- (T2);
            \draw[->,abstractactionedge,densely dotted] (T1) -- (S);
            \draw[->,abstractactionedge,densely dotted] (T2) -- (S);

        \end{tikzpicture}
        }
        
    \caption{Abstract inductive proof graph example, with lemma and action nodes (gray boxes), and associated inductive proof obligations next to each action node. Self-inductive obligations omitted for clarity.}
    \label{fig:abs-ind-proof-with-actions}
\end{figure}

\subsection{Inductive Proof Graph Validity}
We now define a notion of \textit{validity} for an inductive proof graph. That is, we define conditions on when a proof graph can be seen as corresponding to a complete inductive invariant and, correspondingly, when the lemmas of the graph can be determined to be invariants of the underlying system.
\begin{definition}[Local Validity]
    For an inductive proof graph \\ $(V_L \cup V_A,E)$, let the \textit{inductive support set} of an action node $(L,A) \in V_A$ be defined as $Supp_{(L,A)} = \{ \ell \in V_L : (\ell,(L,A)) \in E \}$. We then say that an action node $(L,A)$ is \emph{locally valid} if the following holds:
    \begin{align}
        \left( \wedge_{{\ell \in Supp_{(L,A)}}} \ell \right) \wedge L \wedge A \Rightarrow L'
        \label{eq:local-action-ind-obl}
    \end{align} 
and that a lemma node $L \in V_L$ is \emph{locally valid} if all of its associated action nodes, $\{L\} \times \{A_1,\dots,A_n\}$, are locally valid. We alternately refer to a lemma node that is locally valid as being \textit{discharged}.
\end{definition}
Based on the above local validity definitions, the notion of validity for a full inductive proof graph is then straightforward to define.
\begin{definition}[Inductive Proof Graph Validity]
    An inductive proof graph is \textit{valid} whenever all lemma nodes of the graph are \emph{locally valid}. 
\end{definition}


The validity notion for an inductive proof graph establishes lemmas of such a graph as invariants of the underlying system $M$, since a valid inductive proof graph can be seen to correspond with a complete inductive invariant. We formalize this as follows.



\begin{theorem}
    For a system $M=(I,T)$, if an inductive proof graph $(V_L \cup V_A,E)$ for $M$ is valid, and $I \Rightarrow L$ for every $L \in V_L$, then the conjunction of all lemmas in $V_L$ is an inductive invariant, and all lemmas $L$ are invariants of $M$.
    \label{lemma:ind-graph-invs-inductive}
\end{theorem}
\begin{proof}
    The conjunction of all lemmas in a valid graph must be an inductive invariant, since every lemma's support set exists as a subset of all lemmas in the proof graph, and all lemmas hold on the initial states. Additionally, for any set of predicates, if their conjunction is an invariant of $M$, then each conjunct must be an invariant of $M$, so all such lemmas are necessarily invariants of $M$.
\end{proof}

\subsection{Local Variable Slices}
\label{sec:local-var-slicing}

A valuable feature of the inductive proof graph is that it enables, at each proof node, focus on a smaller subset of state variables relevant for discharging that node.  
%
That is, when considering an action node $(L,A)$, any support lemmas for this node must, to a first approximation, refer only to state variables that appear in either $L$ or $A$. This general idea allows for the computation of a \textit{variable slice} 
at each node, projecting away protocol state variables that are irrelevant for establishing a valid support set for that node. 

Intuitively, the variable slice of an action node $(L,A)$ can be understood as the union of: (1) the set of all variables appearing in the precondition of $A$, (2) the set of all variables appearing in the definition of lemma $L$, (3) for any variables in $L$, the set of all variables upon which the update expressions of those variables depend. 
%
%
More precisely, our slicing computation at each action node is based on the following static analysis of a lemma and action pair $(L,A)$. First, let $\mathcal{V}$ be the set of all state variables in our system, and let $\mathcal{V}'$ refer to the primed, next-state copy of these variables. For an action node $(L,A)$, we have $L \wedge A \Rightarrow L'$ as its initial inductive proof obligation. Like the example protocol from Figure \ref{fig:simple-consensus-spec}, we consider actions to be written in \textit{guarded action} form, so they can be expressed as $A = Pre \wedge Post$, where $Pre$ is a predicate over a set of current state variables, denoted $Vars(Pre) \subseteq \mathcal{V}$, and $Post$ is a conjunction of update expressions of the form $x_i' = f_i(\mathcal{D}_i)$, where $x_i' \in \mathcal{V'}$ and $f_i(\mathcal{D}_i)$ is an expression over a subset of current state variables $\mathcal{D}_i \subseteq \mathcal{V}$. 

\begin{definition}
    For an action $A = Pre \wedge Post$ and variable $x_i' \in \mathcal{V}'$ with update expression $f_i(\mathcal{D}_i)$ in $Post$, we define the \textit{cone of influence} of  $x_i'$, denoted $COI(x_i')$, as the variable set $\mathcal{D}_i$.
    For a set of primed state variables $\mathcal{X}=\{x_1',\dots,x_n'\}$, we define $COI(\mathcal{X})$ simply as $COI(x_1') \cup \dots \cup COI(x_n')$
\end{definition}
Now, if we let $Vars(Pre) \subseteq \mathcal{V}$ and $Vars(L') \subseteq \mathcal{V'}$ be the sets of state variables that appear in the expressions of $L'$ and $Pre$, respectively,
then we can formally define the notion of a slice as follows.
\begin{definition}
    For an action node $(L,A)$, its \textit{variable slice} is the set of state variables
        $Slice(L,A) = Vars(Pre) \cup Vars(L) \cup COI(Vars(L'))$
    \label{def:var-slice-def}
\end{definition}
Based on this, we can now show that a variable slice is a strictly sufficient set of variables to consider when developing a support set for an action node.
\begin{theorem}
    For an action node $(L,A)$, if a valid support set exists, there must exist one whose expressions refer only to variables in $Slice(L,A)$.
\end{theorem}

\begin{proof}
    Without loss of generality, the existence of a support set for $(L,A)$ can be defined as existence of a predicate $Supp$ such that $Supp \wedge L \wedge A \wedge \neg L'$ is unsatisfiable.
    As above, actions are of the form $A = Pre \wedge Post$, where $Post$ is a conjunction of update expressions, $x_i' = f_i(\mathcal{D}_i)$, so this formula can be re-written as
    \begin{align}
        Supp \wedge L \wedge Pre \wedge \neg L'[Post]
        \label{eq:slice-expanded}
    \end{align}
    where $L'[Post]$ represents the expression $L'$ with every $x_i' \in Vars(L')$ substituted with the update expression given by $f_i(\mathcal{D}_i)$. 
    Then, if $L \wedge Pre \wedge \neg L'[Post]$ is satisfiable, and there exists a $Supp$ that makes Formula \ref{eq:slice-expanded} unsatisfiable, then clearly $Supp$ must only refer to variables that appear in $L \wedge Pre \wedge \neg L'[Post]$, which are exactly the set of variables in $Slice(L,A)$.
    

    
\end{proof}


\section{Interactive Safety Verification}
\label{sec:ind-proof-decomp}

Having formalized the inductive proof graph structure in Section \ref{sec:ind-proof-graphs}, we now describe our interactive proof methodology, \textit{inductive proof decomposition}, which is based around incremental construction of an inductive proof graph, with integration of localized counterexample guidance and slicing. We start by providing some additional context and motivation in Section \ref{sec:context-motiv}, followed by a description of our procedure by way of a motivating example in Section \ref{sec:interactive-verif-algo}.

\label{sec:ind-decomp-approach}

\newcommand*{\uniqueleaders}{\emph{UniqueLeaders}}
\newcommand*{\leadersdecide}{\emph{LeadersDecide}}
\newcommand*{\leaderhasquorum}{\emph{LeaderHasQuorum}}
\newcommand*{\noconflictingvalues}{\emph{NoConflictingValues}}

\newcommand{\blue}[1]{\textcolor{blue}{#1}}

\tikzset{%
    safetynode/.style={rounded corners, fill=green!15, draw=blue!50, line width=0.5mm, minimum size=7.5mm},
    unproven/.style={fill=orange!30,label={[label distance=0.0cm]0:\color{orange}{\xmark}}},
    proven/.style={fill=green!15,label={[label distance=0.0cm]0:\color{darkishgreen}{\checkmark}}},
    lemmanode/.style={rounded corners, draw=black, thick, minimum size=7.5mm},
    actionnode/.style={rectangle,draw, fill=green!4,minimum size=3mm},
    slicedvars/.style={font=\small,fill=white,text=black},
}

\subsection{Context and Motivation}
\label{sec:context-motiv}

In a standard interactive safety verification paradigm (e.g., in Ivy \cite{2016padonivy}), the general technique is based on linear, iterative analysis of counterexamples to induction (CTIs). That is, to develop an inductive invariant such as the one shown in Figure~\ref{fig:simple-consensus-spec}, one starts with the target safety property (\emph{NoConflictingValues}) and generates counterexamples to induction, iteratively developing new lemma invariants to rule out these counterexamples until the overall invariant becomes inductive. 

For small protocols and invariants, this basic procedure may be sufficient, but, for large scale verification efforts, its effectiveness breaks down. That is, when systems and their invariants become large, it becomes increasingly difficult to manage and understand the global structure of these inductive invariants as they are being developed, since the interaction between existing lemmas of the invariant candidate and the actions of the system become complex. More concretely, we characterize the issues with existing approaches into the following core themes:
\begin{enumerate}[P1.]
    \item \textbf{CTI Management}: How does one manage the set of CTIs for the current inductive invariant and decide which CTI from this set to analyze?
    \item \textbf{Localization}: Once a CTI is selected, how does one focus only on the lemmas, actions, and state variables relevant to analysis of this CTI?
    \item \textbf{Proof Status and Structure}: As new lemmas are developed, how can the current proof status, structure, and progress be measured?
\end{enumerate}
Our technique, which we describe in the following section, is largely motivated by the fact that, to our knowledge, no existing proof methodologies provide a formal, conceptual framework for addressing the above questions. Thus, existing counterexample-guided inductive invariant development processes are opaque and can be extremely laborious even for relatively experienced protocol designers. In the following section we explain the core ideas and description of our technique and how it aims to address these issues.


\subsection{Interactive Verification Procedure}
\label{sec:interactive-verif-algo}

\input{figures/ind-proof-progression.tex}

We illustrate our interactive verification technique by using it to work through the partial development of an inductive invariant for the \textit{SimpleConsensus} protocol, defined as a symbolic transition system in Figure \ref{fig:simple-consensus-spec}.
This protocol utilizes a simple leader election scheme to select values, and is parameterized on a set of nodes, \emph{Node}, a set of values to be chosen, \emph{Value}, and \emph{Quorum}, a set of intersecting subsets of \emph{Node}. Nodes can vote once for another node to become leader, and once a node $s$ garners a quorum of votes it may become leader and decide a value by setting its local $decided[s]$ variable. The target safety property, \emph{NoConflictingValues}, shown in Figure \ref{fig:simple-consensus-spec}, states that no two differing values can be decided upon by different nodes. 
Also shown there is a complete, monolithic inductive invariant, \emph{Ind}, for establishing this safety property, consisting of 8 lemma conjuncts, along with a subset of definitions for the lemmas in \emph{Ind}.

Figure \ref{fig:ind-proof-graph-progression-init} shows an in-progress inductive proof graph that corresponds to a partially completed inductive invariant containing the first two conjuncts of $Ind$ from Figure \ref{fig:simple-consensus-spec}. 
Our invariant development procedure now starts from this inductive proof graph, from which the possible next steps in our process are clear to assess. 
In particular, it is clear that the $Decide$ and $BecomeLeader$ nodes are unproven (shown in orange with \xmark), meaning that there are outstanding CTIs for the inductive proof obligations of those nodes as explained above. 
Additionally, we can focus on separate CTIs in isolation, since CTIs are associated with specific action nodes. This makes it clear which lemmas and actions these CTIs are relevant to, alleviating the issues of P1 as described above.

To proceed, we then work to extend this graph by developing appropriate support lemmas and associated edges until nodes are discharged (i.e. made valid). 
For example, we may first select $\mathbf{CTI_1}$ to analyze, as shown in Figure \ref{fig:ind-proof-graph-progression-init}. In addition to the localization of counterexamples, the decomposition provided by the proof graph also allows for localized state variable slices to be applied to the CTIs at each action node, and are shown as $V_{slice}$ alongside each action node. 
For example, at the $Decide$ node in Figure \ref{fig:ind-proof-graph-progression-init}, which $\mathbf{CTI_1}$ is associated with, $V_{slice}=\{leader, decided\}$, containing only 2 out of 6 total state variables of the \emph{SimpleConsensus} system. 
These slices, as formalized in Section \ref{sec:local-var-slicing}, significantly reduce the number of state variables to be considered at each node, greatly mitigating the analysis burden on a human user and addressing the issues raised in P2 above.

After analyzing $\mathbf{CTI_1}$ we may develop the {\leadersdecide} lemma, which states that only leaders could have decided values, and rules out $\mathbf{CTI_1}$. {\leadersdecide} is then added as a new incoming support lemma of the \emph{Decide} action node, leading to the proof graph state shown in Figure \ref{fig:ind-proof-graph-progression-2}. 
At this point, we see from the graph status that lemmas {\noconflictingvalues} and {\leadersdecide} are discharged, so we no longer need to consider these lemmas in our reasoning. 
This gives us a useful dynamic measure of proof status and logical structure as we develop new lemmas over time, alleviating the issue of P3 above.

Next, as shown in Figure \ref{fig:ind-proof-graph-progression-2}, we have a remaining  counterexample, $\mathbf{CTI_0}$, associated with the unproven \emph{BecomeLeader} node. 
Here, we again benefit from local slicing, which gives $V_{slice}=\{leader, votes\}$ at this \emph{BecomeLeader} node. 
Analysis of $\mathbf{CTI_0}$ yields the {\leaderhasquorum} support lemma, which is self-inductive, leading us to the proof state in Figure \ref{fig:ind-proof-graph-progression-3}. 
From there, we develop one additional support lemma to rule out $\mathbf{CTI_2}$, giving us the \emph{NodesVoteOnce} support lemma, which is sufficient to discharge the {\uniqueleaders} lemma node, leading us to the proof state shown in Figure \ref{fig:ind-proof-graph-progression-4}.
%
%
We do not show a further progression of the proof process for this graph, but we can see from Figure \ref{fig:ind-proof-graph-progression-4} that this process can be continued in a backwards fashion, starting now from the remaining unproven node \emph{NodesVoteOnce} and its associated action node $RecvVote$. A fully completed proof graph is shown in Figure \ref{fig:simple-consensus-indgraph}, where every node has been discharged i.e., has a valid set of support lemmas.

\begin{figure}[t]
    \centering
    \begin{tikzpicture}[
        every node/.style={scale=0.59}
    ]
        \node[safetynode,proven] (safety) at (0,0) {\emph{NoConflictingValues}};
        \node[actionnode,proven,label={[slicedvars]-40:$V_{slice}$=$\{leader,decided\}$}] (decide_S) [above=0.4cm of safety] {\emph{Decide}};

        \node[lemmanode,proven] (leadersDecide) [above left=0.3cm of decide_S] {\emph{LeadersDecide}};

        \node[lemmanode,proven] (uniqueLeaders) [above right=0.3cm of decide_S] {\emph{UniqueLeaders}};
        \node[actionnode,proven,label={[slicedvars]-40:$V_{slice}$=$\{leader,votes\}$}] (becomeLeader1) [above=0.4cm of uniqueLeaders] {\emph{BecomeLeader}};
        \node[lemmanode,proven] (LeaderHasQuorum) [above right=0.3cm of becomeLeader1] {\emph{LeaderHasQuorum}};
        \node[lemmanode,proven] (nodesVoteOnce) [above left=0.3cm of becomeLeader1] {\emph{NodesVoteOnce}};

        \node[actionnode,proven,label={[slicedvars]-40:$V_{slice}$=$\{voteMsg,votes\}$}] (recvVoteAction) [above=0.4cm of nodesVoteOnce] {\emph{RecvVote}};
        \node[lemmanode,proven] (voteImpliesVoteMsg) [above right=0.3cm of recvVoteAction] {\emph{VoteRecvdImpliesVoteMsg}};
        \node[lemmanode,proven] (voteMsgsUnique) [above left=0.3cm of recvVoteAction] {\emph{VoteMsgsUnique}};
        
        \node[actionnode,proven,label={[slicedvars]-40:$V_{slice}$=$\{voteMsg,voteRequestMsg,voted\}$}] (sendVoteAction) [above=0.4cm of voteMsgsUnique] {\emph{SendVote}};

        \node[lemmanode,proven] (voteMsgImpliesVoted) [above=0.3cm of sendVoteAction] {\emph{VoteMsgImpliesVoted}};

        \draw[->,densely dotted] (decide_S) -> (safety);
        \draw[->,thick] (uniqueLeaders) -> (decide_S);
        \draw[->,thick] (leadersDecide) -> (decide_S);
        \draw[->,densely dotted] (becomeLeader1) -> (uniqueLeaders);
        \draw[->,thick] (LeaderHasQuorum) -> (becomeLeader1);
        \draw[->,thick] (nodesVoteOnce) -> (becomeLeader1);
        \draw[->,densely dotted] (recvVoteAction) -> (nodesVoteOnce);
        \draw[->,thick] (voteImpliesVoteMsg) -> (recvVoteAction);
        \draw[->,thick] (voteMsgsUnique) -> (recvVoteAction);
        \draw[->,densely dotted] (sendVoteAction) -> (voteMsgsUnique);
        \draw[->,thick] (voteMsgImpliesVoted) -> (sendVoteAction);
    \end{tikzpicture}  
    \caption{Complete inductive proof graph for \textit{SimpleConsensus} and safety property \emph{NoConflictingValues}, with variable slice annotations.}
    \label{fig:simple-consensus-indgraph}
\end{figure}

Even on this relatively small protocol and invariant, this example demonstrates the value of our interactive, compositional proof methodology. Our decomposition based on the inductive proof graph structure makes explicit the relationship between lemmas and protocol actions as the inductive invariant is being built, and, enabled by this structure, we are able to localize the analysis of CTIs to specific nodes of this graph, along with the benefits of local slicing.
We describe this general procedure slightly more formally in Procedure \ref{fig:interactive-verif-algo}. That is, our high level interactive safety verification procedure, for proving that $S$ is an invariant of transition system $M=(I,T)$ by finding an inductive invariant $Ind$, centers around the incremental construction of an inductive proof graph, working backwards from $S$ as the target invariant.




\makeatletter
\renewcommand{\ALG@name}{Procedure}
\makeatother
\begin{algorithm}[t]
    \caption{Abstract steps of our interactive verification procedure.}
    \label{fig:interactive-verif-algo}
    \footnotesize
    \input{figures/interactive-verif-algo.tex}
\end{algorithm}

\section{Empirical Evaluation}
\label{sec:evaluation}

To evaluate our technique, we used it to develop inductive invariants for several distributed protocols of varying complexity, including development of an inductive invariant for a large scale specification of the Raft \cite{raftpaper} consensus protocol, the first inductive invariant effort for a Raft protocol specification of this complexity with this degree of automation. 

An artifact containing all of our source code and instructions
for reproducing our evaluation results can be found at \cite{nfm26artifact}. A
public, open-source version of our tool is also available at \cite{scimitar}.

\paragraph{Implementation and Benchmarks}

\newcommand{\toolname}{\textsc{Scimitar}}
\newcommand{\apalacheversion}{0.44.0}
\newcommand{\tlcversion}{2.15}

We implemented our inductive proof decomposition technique in a tool, \toolname{}, that 
provides a graphical user interface for the interactive development of inductive proof graphs. \toolname{} is implemented in Python and accepts systems specified in the \tlaplus{} specification language \cite{lamport2002specifying}. The tool allows a user to view a current inductive proof graph, with visualizations representing the state of each lemma and action node. A user can then choose to focus on a particular action node, and a subset of CTIs associated with that node are presented. Slicing is applied automatically to the presented CTIs based on an automatic static analysis of the protocol specification, and the interface also provides a way to for a user to dynamically add support edges and new lemmas to the graph.

Internally, \toolname{} uses a combination of the TLC model checker \cite{tlcmodelchecker} and the Apalache symbolic model checker \cite{2019tlamadesymbolic} for checking inductive proof obligations and generating CTIs. Note that TLC is an explicit state model checker, so cannot produce full proofs of inductive invariants for protocols of arbitrary size, but can generate CTIs for finite protocol instances using a randomized search technique \cite{Lamport2018UsingTT}. We used finite parameter bounds for counterexample generation with TLC and Apalache, and then also use the \tlaplus{} proof system (TLAPS) \cite{cousineau2012tla} to fully validate the final inductive proof graphs.
In our experience, TLC can, in some cases, be more efficient than Apalache at generating CTIs, so we found both tools complementary in our implementation. All experiments and proof checking results below were collected using Apalache version \apalacheversion{} and TLC version \tlcversion{} running on a 2023 M3 Macbook Pro. 


\newcommand{\twophaseparamsrms}{\{s_1,s_2,s_3,s_4\}}

\newcommand{\abstractraftparamservers}{\{s_1,s_2,s_3,s_4\}}
\newcommand{\abstractraftparammaxlen}{3}
\newcommand{\abstractraftparammaxterms}{3}

\newcommand{\asyncraftparamservers}{\{s_1,s_2,s_3\}}
\newcommand{\asyncraftparammaxlen}{3}
\newcommand{\asyncraftparammaxterms}{3}

\newcommand{\zabparamservers}{\{s_1,s_2,s_3\}}
\newcommand{\zabparammaxepoch}{2}
\newcommand{\zabparammaxhistlen}{1}

We used our tool to develop inductive invariants for establishing core safety properties of 5 distributed protocol specifications. These protocols are all specified in \tlaplus{}, and summarized in Table \ref{fig:protocol-table}, along with various specification and proof statistics.  
%
Of the protocols tested, we consider the following 3 to be of medium complexity:
\begin{itemize}
    \item \emph{SimpleConsensus}: An abstract consensus protocol where nodes elect a leader which then decides on a value, as shown in Figure \ref{fig:simple-consensus-spec}.
    \item \emph{TwoPhase}: A specification of the two-phase commit protocol \cite{lampson2pc}. The safety property is \textit{TCConsistent}, stating that two resource managers cannot have conflicting commit or abort decisions. 
    \item \emph{AbstractRaft}: A high level specification of the Raft consensus protocol \cite{raftpaper}, abstracting away message passing. The safety property is \mbox{\emph{StateMachineSafety}}, stating that committed entries must be consistent across nodes. 
\end{itemize}
The 2 additional protocols are of significantly larger complexity:
\begin{itemize}
    \item \textit{Bakery}: A specification of Lamport's Bakery algorithm \cite{1974lamportbakery} for mutual exclusion. The safety property is \textit{MutualExclusion}, stating two processes cannot be in a critical section simultaneously.
    \item \textit{AsyncRaft}: An industrial scale specification of the Raft consensus protocol \cite{raftpaper}, based on \cite{OngaroDissertation2014,Vanlightly2023}, and which models asynchronous message passing between nodes and fine-grained local state. The safety property is \mbox{\textit{NoLogDivergence}}, which asserts that committed indices across nodes are consistent.
\end{itemize}
The 2 large protocol specifications are both of a complexity significantly greater than those tested in recent automated invariant inference techniques, so are more relevant for evaluating our interactive verification techniques. 

\newcommand{\bmdir}{/Users/willyschultz/Dropbox/PhD/Research/inductive-invariant-inference/endive/benchmarks}

\newcommand*{\simpleconsensusnumlemmas}{9}
\newcommand*{\simpleconsensuslocspec}{77}
\newcommand*{\simpleconsensuslocproof}{51}
\newcommand*{\simpleconsensuschecktime}{72}
\newcommand*{\twophasechecktime}{184}
\newcommand*{\abstractraftchecktime}{969}
\newcommand*{\asyncraftchecktime}{6592}

\newcommand*{\Zabchecktime}{1101}

\input{diagrams/consensus_epr_proofstats.tex}
\input{diagrams/TwoPhase_proofstats.tex}
\input{diagrams/AbstractRaft_proofstats.tex}
\input{diagrams/Bakery_proofstats.tex}
\input{diagrams/AsyncRaft_proofstats.tex}

\begin{table}
    \small
    \centering
    \begin{tabular}{l|c|c|c|c|c|c}
        \textbf{Protocol} & Variables & Actions &  LOC (spec/pf) & Lemmas & Median Slice Size \\
        \hline
        SimpleConsensus & \consensuseprnumstatevars & \consensuseprnumactions & \makebox[0.4cm][c]{\consensuseprspecloc} / \makebox[0.4cm][r]{\consensuseprproofloc} & \consensuseprnumlemmas & \consensuseprmedianslicesize \, (\consensuseprmedianslicepct)  \\
        \hline
        TwoPhase & \TwoPhasenumstatevars & \TwoPhasenumactions & \makebox[0.4cm][c]{\TwoPhasespecloc} / \makebox[0.4cm][r]{\TwoPhaseproofloc} & \TwoPhasenumlemmas & \TwoPhasemedianslicesize \, (\TwoPhasemedianslicepct)  \\
        \hline
        AbstractRaft & \AbstractRaftnumstatevars & \AbstractRaftnumactions & \makebox[0.4cm][c]{\AbstractRaftspecloc} / \makebox[0.4cm][r]{\AbstractRaftproofloc} & 10 & \AbstractRaftmedianslicesize \, (\AbstractRaftmedianslicepct)   \\
        \hline 
        \hline
        Bakery & \Bakerynumstatevars & \Bakerynumactions & \makebox[0.4cm][c]{\Bakeryspecloc} / \makebox[0.4cm][r]{\Bakeryproofloc} & \Bakerynumlemmas & \Bakerymedianslicesize \, (\Bakerymedianslicepct)  \\
        \hline
        AsyncRaft & 12 & \AsyncRaftnumactions & \makebox[0.4cm][c]{\AsyncRaftspecloc} / \makebox[0.4cm][r]{\AsyncRaftproofloc} & \AsyncRaftnumlemmas & \AsyncRaftmedianslicesize \, (\AsyncRaftmedianslicepct)  \\
        \hline
    \end{tabular}
    \caption{Protocol benchmarks and associated metrics. \textit{LOC} is the number of lines of \tlaplus{} code for defining the specification (spec) and inductive proof (pf), respectively. \textit{Median slice} gives the median size of all variable slices in the complete inductive proof graph, and its proportion of the total state variables.
    }
    \label{fig:protocol-table}
\end{table}


\subsection{Results and Discussion}

Table \ref{fig:protocol-table} shows various statistics about the protocols we tested, including the number of state variables, number of actions, lines of code (LOC) in the \tlaplus{} protocol specifications and proofs, number of lemmas in each proof graph, and the size of variable slices. 
We discuss the structure of the developed proof graphs and qualitative aspects of our experience developing these proofs in more detail below.

\subsubsection*{Medium Protocols}

Visualizations of the completed inductive proof graphs for \textit{AbstractRaft} and \textit{TwoPhase}, respectively, can be seen in Figures \ref{fig:twophase-proof-struct} and \ref{fig:absraft-proof-struct}, and the proof graph for \textit{SimpleConsensus} was shown previously, in Figure \ref{fig:simple-consensus-indgraph}. The full graphs can be viewed and explored using our interactive tool.

Though these proof graphs are for smaller protocols, they generally confirm our intuition that these graphs are useful for understanding the structure of an inductive proof and guiding its development. Even for protocols of this size, we found that maintenance of this structure coupled with automatic counterexample slicing is a significant aid, and allowed us to develop these inductive invariants with much more ease and efficiency. In \textit{SimpleConsensus}, for example, the median slice proportion is $0.33$, indicating that most slices present a significantly reduced amount of information, and the graph admits a clear tree-like decomposition.
Our experience for \textit{TwoPhase} was similar, also benefitting from slicing with a median slice proportion of 0.5.
%

The graph for \textit{AbstractRaft} similarly admits a relatively tree-like decomposition, and also provides insight on how lemmas of the protocol relate to each other. For example, we can observe an \textit{induction cycle} related to establishment of key properties about committed log entries. This cycle flows from \emph{CommittedEntryIsOnQuorum} to \emph{LeaderCompleteness} via action \emph{BecomeLeader}, then to \emph{LaterLogsHaveEarlierCommitted} via action \emph{ClientRequest}, then back to \emph{CommittedEntryIsOnQuorum} via \emph{RollbackEntries}. Generally, during development of these inductive proofs, we found this structure helpful to guide our reasoning as it makes the current logical structure of the developed proof explicit.




\tikzset{%
    prooflemmaedge/.style={line width=1.5pt},
    proofactionedge/.style={line width=1.5pt,dotted},
    proofactionnode/.style={fill=lightgray!30,scale=1.15,font=\small,rounded corners},
    edgehighlight/.style={line width=2.0pt,black,preaction={draw, line width=2.5mm, blue!20, -}},
    broken/.style={fill=orange!40,label={[label distance=0.0cm]0:\color{orange}{\xmark}}}
}


\tikzset{%
    safetynode/.style={rounded corners,fill=green!20, draw=blue!50, line width=0.5mm, minimum size=7.5mm},
    safetynode_detailed/.style={rounded corners, scale=1.9,fill=green!20, draw=blue!50, line width=0.5mm, minimum size=7.5mm},
    es_support/.style={rounded corners, fill=lightblue!40, draw=black, line width=0.9mm, minimum size=7.5mm},
    keylemmanode/.style={rounded corners, fill=blue!20, draw=blue!50, line width=0.9mm, minimum size=7.5mm,font=\large},
    unproven/.style={fill=orange!30,label={[label distance=0.0cm]0:\color{orange}{\xmark}}},
    failed/.style={fill=red!30,label={[label distance=0.0cm]0:\color{red}{\xmark}}},
    proven/.style={fill=green!20,label={[label distance=0.0cm]0:\color{darkishgreen}{\checkmark}}},
    lemmanode/.style={rounded corners, draw=black, thick, minimum size=7.5mm,font=\large},
    lemmanode_detailed/.style={rounded corners, draw=black, thick, minimum size=7.5mm,font=\large,scale=1.9,line width=0.85mm},
    lemmanode_detailed_raft/.style={rounded corners, draw=black, thick, minimum size=7.5mm,font=\large,scale=1.9,line width=0.85mm},
    actionnode/.style={rectangle,draw, fill=green!4,minimum size=3mm},
    actionnode_detailed/.style={rectangle,draw, fill=green!20,minimum size=3mm,font=\huge,scale=1.0},
    slicedvars/.style={fill=none,text=black},
    slicedvars_detailed/.style={fill=none,text=black,scale=1.2,font=\large},
    actionedge_detailed/.style={line width=0.55mm},
    cycleedge/.style={draw=blue}
}

\begin{figure}[t]
        \centering
        \scalebox{.33}{\input{diagrams/TwoPhase.proof.dot_nolabels.tex}}
    \caption{\emph{TwoPhase} inductive proof graph.}
    \label{fig:twophase-proof-struct}
\end{figure}

\begin{figure}[t]
        \centering
        \scalebox{.28}{\input{diagrams/AbstractRaft.proof.dot.tex}}
    \caption{\emph{AbstractRaft} inductive proof graph.}
    \label{fig:absraft-proof-struct}
\end{figure}





\subsubsection*{Large Protocols}

\newcommand{\AsyncRaftFigPreamble}{
}
\newcommand{\AsyncRaftWithProofStatusFigPreamble}{
    \clip (0,1) rectangle + (17,18);
}

\makeatletter
\newcommand{\Larger}{\@setfontsize\Larger{14}{16}}
\makeatother

\begin{figure}[h]
    \centering
    \scalebox{.2}{\input{diagrams/AsyncRaft.proof.dot_nolabels.tex}}
    \caption{\emph{AsyncRaft} inductive proof graph structure, with excerpted action nodes and lemmas shown with detailed names, and induction cycle highlighted in blue. }
    \label{fig:async-raft-proof-struct}
\end{figure}

As an exploration of our technique for larger scale proof efforts, we applied it first to \textit{AsyncRaft}, a specification of Raft at a considerably lower level of abstraction e.g., it includes asynchronous message passing and fine-grained local state, akin to the detail level provided in Raft's original \tlaplus{} specification \cite{OngaroDissertation2014}. We are also aware of no prior inductive invariant that existed for a \tlaplus{} specification of Raft at this level of complexity. Figure \ref{fig:async-raft-proof-struct} shows a visualization of our completed inductive proof graph for \textit{AsyncRaft}. 


Overall, we found our technique effective at making the proof process efficient and understandable for a human verifier, by providing a formal structure to the large scale proof and accelerating CTI analysis by slicing and localized reasoning. We were able to develop such an inductive invariant in approximately 3 human weeks of effort, which we found to be a productive pace for an invariant of this size and protocol of this complexity. Other verification efforts of this type note, for example, an inductive invariant development burden of 1-2 human months \cite{schultz2021formal}, even for a protocol with a smaller invariant than \textit{AsyncRaft}. We found that the decomposition provided by our technique enabled effective local reasoning in many cases. For example, comparing the slices near \emph{LeaderMatchIndexValid} and \emph{LeaderMatchIndexBound} nodes to \emph{QuorumsSafeAtTerms}, in a relatively distinct sub-component of the graph, we can see that the slice sets refer to relatively disjoint subsets of variables (out of 12 total variables), supporting our hypothesis that the compositional structure of the proof graph allows for localized reasoning on different aspects of the protocol.


In addition to the efficiency of invariant development, we also found value in the developed proof artifact beyond establishing correctness. 
%
For example, during the development of the proof, we found it helpful to observe various patterns that arise in the proof graph structure. 
In particular, we observed the various types of \textit{induction cycles} that arise in this graph, especially those that arise in many places due to the message passing nature of this protocol. For example, we can observe the cycle highlighted in Figure \ref{fig:async-raft-proof-struct} where \emph{LogMatching} serves as a support lemma of \emph{LogMatchingInAppendEntriesMsgs} via the \emph{AppendEntries} action, and \emph{LogMatchingInAppendEntriesMsgs} then serves as a support lemma of \emph{LogMatching} via the \emph{AppendEntriesResponse} action, forming this induction cycle. These cycles would be difficult to observe in a standard inductive invariant, but are apparent in our proof graph structure, and represent a common pattern where invariants about protocol state must hold on both local state and also on the state of messages that are sent over the network.
%
We found a similar experience with \textit{Bakery} proof development, but omit its full proof graph for sake of space, showing its summarized proof statistics in Table \ref{fig:protocol-table}.

\section{Related Work}

\label{sec:related-work}


There are two recent works that are most similar to ours in scope and approach, namely, the Ivy system \cite{2016padonivy} and the work on exploiting modularity for decidability presented in \cite{taube2018modularity}. 
Our approach bears similarities to these other two verification techniques, but there are a number of key differences in terms of goals and methodologies. 

One main focus of Ivy is on the modeling language, with a goal of making it easy to represent systems in a decidable fragment of first order logic, so as to ensure verification conditions always provide some concrete feedback in the form of counterexamples. They also discuss an interactive approach for generalization from counterexamples, that has similarities to the UPDR approach used in extensions of IC3/PDR \cite{2017updr}. In contrast, our work is primarily focused on different concerns e.g., we focus on compositionality as a means to provide an efficient and scalable proof methodology, and as a means to produce a more interpretable proof artifact, in addition to allowing for localized counterexample reasoning. We also view decidable modeling as an orthogonal component of the verification process that could be complementary to our approach.

The goals of the work in \cite{taube2018modularity} are more similar to our own, in that they aim to use a particular type of compositional reasoning to exploit decidable subproblems when possible. This is perhaps closest to our approach in that it tries to exploit decomposition in the verification and proof process, but the decomposition notions and the way they are used are somewhat different between our work and theirs. Our goal is to define a compositional structure that is integrated into the counterexample guidance process, while also producing a single, inductive proof artifact upon completion. 
In future it would be interesting, however, to consider whether our notion of decomposition based on the inductive proof graph structure admits a similar kind of ``decidable subproblem" property that is exploited in \cite{taube2018modularity}.




Our techniques also bear similarities to prior approaches for the proofs and analysis of concurrent programs, namely that of \textit{inductive data flow graphs} \cite{Farzan2015InductiveDF}. That work, however, is focused on the verification of multi-process concurrent programs, and did not generalize the notions to a distributed setting. The procedures for verification and counterexample analysis are also different between our approach and theirs. Our compositional approach also bears similarity to recent work on \textit{recomposition} techniques for verification \cite{2024recomposition}.
Generally, our slicing technique is similar to a cone-of-influence reduction \cite{2011COI}, as well as other \textit{program slicing} techniques \cite{Tip1994ASO}. It also shares some concepts with other path-based program analysis techniques that incorporate slicing techniques \cite{jaffar2012path,jhala2005path}. In our case, however, we apply it at the level of a single protocol action and target lemma, particularly for the purpose of CTI state projection.

With respect to fully automated invariant inference, there are several recently published techniques that attempt to solve this problem for distributed protocols, including IC3PO \cite{2021ic3posymmetry}, SWISS \cite{2021swisshance} and DuoAI \cite{YaoTGN22}. These tools, however, provide little feedback when they fail on a given problem, and the large scale protocols we presented in this paper, are of a complexity considerably higher than what modern tools in this area can solve.


%

\section{Conclusions and Future Work}
\label{sec:conclusions}

We presented \textit{inductive proof decomposition}, a new methodology for human-guided development of inductive invariants for large-scale protocol safety verification. In future, we are interested in exploring approaches enabled by this technique and proof structure e.g., integrating more automation by applying local syntax-guided synthesis techniques to proof graph nodes. We are also interested in understanding how these proof graph structures might be useful as a part of other automated model checking engines, and in understanding the empirical structure of these proof graphs for additional real world protocols.

\bibliographystyle{acm}
\bibliography{references}

\end{document}

%% file: figures/ind-proof-progression.tex
\begin{figure}[t]
    \begin{subfigure}[b]{0.52\textwidth}
        \centering
        \begin{tikzpicture}[
            every node/.style={scale=0.6}
        ]
        \node[safetynode,unproven] (safety) at (0,0) {\emph{NoConflictingValues}};
        \node[actionnode,unproven,label={[slicedvars]-40:$V_{slice}$=$\{leader,decided\}$}] (decide_S) [above=0.5cm of safety] {\emph{Decide}};


        \node[lemmanode,unproven] (uniqueLeaders) [above right=0.2cm of decide_S] {\emph{UniqueLeaders}};
        \node[actionnode,unproven,label={[slicedvars]-40:$V_{slice}$=$\{leader,votes\}$}] (becomeLeader1) [above=0.5cm of uniqueLeaders] {\emph{BecomeLeader}};



        %


        %
        %

        \begingroup
        \addtolength{\jot}{-0.5em}

        \node[draw,rounded corners] (CTI1) at (-1.65,1.65) {
        $\begin{aligned}
            &\mathbf{CTI_1} \, (Sliced)\\
            &\, \text{leader} = (n_1 \mapsto \true,  n_2 \mapsto \false) \\
            &\, \text{decided} = (n_1 \mapsto \{\}, n_2 \mapsto \{v_1\}) \\
        \end{aligned}$
        };



        \node[draw,rounded corners] (CTI2) at (-1.65,3.05) {
            $\begin{aligned}
                &\mathbf{CTI_{0}} \, (Sliced)\\
                &\, \text{leader} = (n_1 \mapsto \true, n_2 \mapsto \false)  \\
                &\, \text{votes} = (n_1 \mapsto \{\}, n_2 \mapsto \{n_1, n_2\})
            \end{aligned}$
            };
        \endgroup




        \draw [gray] (becomeLeader1) -- (CTI2);
        \draw [gray] (decide_S) -- (CTI1);

        \draw[->,densely dotted] (decide_S) -> (safety);
        \draw[->,thick] (uniqueLeaders) -> (decide_S);
        \draw[->,densely dotted] (becomeLeader1) -> (uniqueLeaders);
        \end{tikzpicture}  
        \caption{Proof step 1.}
        \label{fig:ind-proof-graph-progression-init}
    \end{subfigure}
    \vspace{0.35cm}
    \begin{subfigure}[b]{0.46\textwidth}
        \centering
        \begin{tikzpicture}[
            every node/.style={scale=0.6}
        ]
        \node[safetynode,proven] (safety) at (0,0) {\emph{NoConflictingValues}};
        \node[actionnode,proven,label={[slicedvars]-40:$V_{slice}$=$\{leader,decided\}$}] (decide_S) [above=0.5cm of safety] {\emph{Decide}};

        \node[lemmanode,proven] (leadersDecide) [above left=0.2cm of decide_S] {\emph{LeadersDecide}};

        \node[lemmanode,unproven] (uniqueLeaders) [above right=0.2cm of decide_S] {\emph{UniqueLeaders}};
        \node[actionnode,unproven,label={[slicedvars]-40:$V_{slice}$=$\{leader,votes\}$}] (becomeLeader1) [above=0.5cm of uniqueLeaders] {\emph{BecomeLeader}};

        %


        %
        %

        \begingroup
        \addtolength{\jot}{-0.5em}

        \node[draw,rounded corners] (CTI0) at (-1.55,3.05) {
            $\begin{aligned}
                &\mathbf{CTI_{0}} \, (Sliced)\\
                &\, \text{leader} = (n_1 \mapsto \true, n_2 \mapsto \false)  \\
                &\, \text{votes} = (n_1 \mapsto \{\}, n_2 \mapsto \{n_1, n_2\})
            \end{aligned}$
            };
        \endgroup





        \draw [gray] (becomeLeader1) -- (CTI0);

        \draw[->,densely dotted] (decide_S) -> (safety);
        \draw[->,thick] (uniqueLeaders) -> (decide_S);
        \draw[->,thick] (leadersDecide) -> (decide_S);
        \draw[->,densely dotted] (becomeLeader1) -> (uniqueLeaders);
        \end{tikzpicture}  
        \caption{Proof step 2.}
        \label{fig:ind-proof-graph-progression-2}
    \end{subfigure}
    \begin{subfigure}[b]{0.45\textwidth}
        \centering
        \begin{tikzpicture}[
            every node/.style={scale=0.6}
        ]
        \node[safetynode,proven] (safety) at (0,0) {\emph{NoConflictingValues}};
        \node[actionnode,proven,label={[slicedvars]-40:$V_{slice}$=$\{leader,decided\}$}] (decide_S) [above=0.5cm of safety] {\emph{Decide}};

        \node[lemmanode,proven] (leadersDecide) [above left=0.2cm of decide_S] {\emph{LeadersDecide}};

        \node[lemmanode,unproven] (uniqueLeaders) [above right=0.2cm of decide_S] {\emph{UniqueLeaders}};
        \node[actionnode,unproven,label={[slicedvars]-40:$V_{slice}$=$\{leader,votes\}$}] (becomeLeader1) [above=0.5cm of uniqueLeaders] {\emph{BecomeLeader}};

        \node[lemmanode,proven] (LeaderHasQuorum) [above right=0.2cm and -0.7cm of becomeLeader1] {\emph{LeaderHasQuorum}};


        %


        %
        %

        \begingroup
        \addtolength{\jot}{-0.5em}


        \node[draw,rounded corners] (CTI2) at (-0.8,3.15) {
        $\begin{aligned}
            &\mathbf{CTI_{2}} \, (Sliced)\\
            &\, \text{leader} = (n_1 \mapsto \true, n_2 \mapsto \false)  \\
            &\, \text{votes} = (n_1 \mapsto \{n_1, n_2\}, n_2 \mapsto \{n_1, n_2\})
        \end{aligned}$
        };
        \endgroup




        \draw [gray] (becomeLeader1) -- (CTI2);

        \draw[->,densely dotted] (decide_S) -> (safety);
        \draw[->,thick] (uniqueLeaders) -> (decide_S);
        \draw[->,thick] (leadersDecide) -> (decide_S);
        \draw[->,densely dotted] (becomeLeader1) -> (uniqueLeaders);
        \draw[->,thick] (LeaderHasQuorum) -> (becomeLeader1);
        \end{tikzpicture}  
        \caption{Proof step 3.}
        \label{fig:ind-proof-graph-progression-3}
    \end{subfigure}
    \hfill
    \begin{subfigure}[b]{0.45\textwidth}
        \centering
        \begin{tikzpicture}[
            every node/.style={scale=0.6}
        ]
        \node[safetynode,proven] (safety) at (0,0) {\emph{NoConflictingValues}};
        \node[actionnode,proven,label={[slicedvars]-40:$V_{slice}$=$\{leader,decided\}$}] (decide_S) [above=0.5cm of safety] {\emph{Decide}};

        \node[lemmanode,proven] (leadersDecide) [above left=0.2cm of decide_S] {\emph{LeadersDecide}};

        \node[lemmanode,proven] (uniqueLeaders) [above right=0.2cm of decide_S] {\emph{UniqueLeaders}};
        \node[actionnode,proven,label={[slicedvars]-40:$V_{slice}$=$\{leader,votes\}$}] (becomeLeader1) [above=0.5cm of uniqueLeaders] {\emph{BecomeLeader}};
        \node[lemmanode,proven] (LeaderHasQuorum) [above right=0.2cm and -0.7cm of becomeLeader1] {\emph{LeaderHasQuorum}};
        \node[lemmanode,unproven] (nodesVoteOnce) [above left=0.2cm and -0.2cm of becomeLeader1] {\emph{NodesVoteOnce}};

        \node[actionnode,unproven,label={[slicedvars]-40:$V_{slice}$=$\{voteMsg,votes\}$}] (recvVoteAction) [above=0.4cm of nodesVoteOnce] {\emph{RecvVote}};
        %


        %
        %



        \draw[->,densely dotted] (decide_S) -> (safety);
        \draw[->,thick] (uniqueLeaders) -> (decide_S);
        \draw[->,thick] (leadersDecide) -> (decide_S);
        \draw[->,densely dotted] (becomeLeader1) -> (uniqueLeaders);
        \draw[->,thick] (LeaderHasQuorum) -> (becomeLeader1);
        \draw[->,thick] (nodesVoteOnce) -> (becomeLeader1);
        \draw[->,densely dotted] (recvVoteAction) -> (nodesVoteOnce);
        \end{tikzpicture}  
        \caption{Proof step 4.}
        \label{fig:ind-proof-graph-progression-4}
    \end{subfigure}
    \caption{Example progression of inductive proof graph development for invariant of Fig. \ref{fig:simple-consensus-spec}. Nodes in orange(\xmark{}) are those with remaining inductive proof obligations to be discharged, and those in green(\checkmark{}) have all obligations discharged. CTI pre-states are shown as annotations associated with relevant action nodes.}
    \label{fig:ind-proof-graph-progression}
\end{figure}

%% file: figures/interactive-verif-algo.tex
\algnewcommand{\algorithmicgoto}{\textbf{goto}}%
\algnewcommand{\Goto}[1]{\algorithmicgoto~Line \ref{#1}}%
\begin{algorithmic}[1]
    \State \textbf{Inputs}: Transition system $M$, invariant $S$.
    \State \textbf{Initialize}: $V_L \gets \{S\}$; $V_A \gets \{S\} \times \{A_1,\dots,A_k\}$; $E \gets \emptyset$; $G \gets (V_L \cup V_A,E)$
    \Procedure{InductiveProofDecomposition}{}
    \If{all lemmas $V_L$ of $G$ are locally valid} \label{line:ifbegin}
        \State \Return $G$, a valid inductive proof graph.
        \Else
        \State Choose some $(L,A) \in V_A$ such that $(L,A)$ is not locally valid. 
        \State Analyze CTI $X$ at node $(L,A)$, develop lemma $L_{new}$ eliminating $X$ 
        \State Update $G$ as: \label{line-lnew}
        \State $V_L \gets V_L \cup \{ L_{new} \}$
        \State $V_A \gets V_A \cup (\{ L_{new} \} \times \{A_1,\dots,A_k\})$ 
        \State $E \gets E \cup \{ (L_{new},(L,A)) \}$  
        \State \Goto{line:ifbegin}.
    \EndIf
    %
    \EndProcedure
\end{algorithmic}

%% file: diagrams/consensus_epr_proofstats.tex
\newcommand*{\consensuseprnumstatevars}{6}
\newcommand*{\consensuseprmeanindegree}{0}
\newcommand*{\consensuseprmedianslicesize}{2}
\newcommand*{\consensuseprmedianindegree}{1}
\newcommand*{\consensuseprnumlemmas}{8}
\newcommand*{\consensuseprnumactions}{5}
\newcommand*{\consensuseprspecloc}{111}
\newcommand*{\consensuseprproofloc}{18}
\newcommand*{\consensuseprmedianslicepct}{0.33}

%% file: diagrams/TwoPhase_proofstats.tex
\newcommand*{\TwoPhasenumstatevars}{6}
\newcommand*{\TwoPhasemeanindegree}{1}
\newcommand*{\TwoPhasemedianslicesize}{3}
\newcommand*{\TwoPhasemedianindegree}{1}
\newcommand*{\TwoPhasenumlemmas}{16}
\newcommand*{\TwoPhasenumactions}{7}
\newcommand*{\TwoPhasespecloc}{192}
\newcommand*{\TwoPhaseproofloc}{140}
\newcommand*{\TwoPhasemedianslicepct}{0.50}

%% file: diagrams/AbstractRaft_proofstats.tex
\newcommand*{\AbstractRaftnumstatevars}{4}
\newcommand*{\AbstractRaftmeanindegree}{1}
\newcommand*{\AbstractRaftmedianslicesize}{3}
\newcommand*{\AbstractRaftmedianindegree}{1}
\newcommand*{\AbstractRaftnumlemmas}{12}
\newcommand*{\AbstractRaftnumactions}{6}
\newcommand*{\AbstractRaftspecloc}{220}
\newcommand*{\AbstractRaftproofloc}{138}
\newcommand*{\AbstractRaftmedianslicepct}{0.75}

%% file: diagrams/Bakery_proofstats.tex
\newcommand*{\Bakerynumstatevars}{6}
\newcommand*{\Bakerymeanindegree}{1}
\newcommand*{\Bakerymedianslicesize}{4}
\newcommand*{\Bakerymedianindegree}{2}
\newcommand*{\Bakerynumlemmas}{27}
\newcommand*{\Bakerynumactions}{14}
\newcommand*{\Bakeryspecloc}{283}
\newcommand*{\Bakeryproofloc}{116}
\newcommand*{\Bakerymedianslicepct}{0.67}

%% file: diagrams/AsyncRaft_proofstats.tex
\newcommand*{\AsyncRaftnumstatevars}{12}
\newcommand*{\AsyncRaftmeanindegree}{2}
\newcommand*{\AsyncRaftmedianslicesize}{5}
\newcommand*{\AsyncRaftmedianindegree}{2}
\newcommand*{\AsyncRaftnumlemmas}{39}
\newcommand*{\AsyncRaftnumactions}{11}
\newcommand*{\AsyncRaftspecloc}{638}
\newcommand*{\AsyncRaftproofloc}{920}
\newcommand*{\AsyncRaftmedianslicepct}{0.42}

%% file: diagrams/TwoPhase.proof.dot_nolabels.tex
\begin{tikzpicture}[>=latex,line join=bevel,]

    \Large
\node (H_TCConsistent) at (346.37bp,10.579bp) [draw=black,fill=green!50,rectangle] {\huge\emph{TCConsistent}};
\node (H_TCConsistent_RMRcvAbortMsgAction) at (195.37bp,65.107bp) [draw,rectangle,fill=lightgray,proofactionnode] {$RMRcvAbortMsg$};
\node (H_TCConsistent_RMChooseToAbortAction) at (346.37bp,65.107bp) [draw,rectangle,fill=lightgray,proofactionnode] {$RMChooseAbort$};
\node (H_TCConsistent_RMRcvCommitMsgAction) at (586.37bp,121.76bp) [draw,rectangle,fill=lightgray,proofactionnode] {$RMRcvCommitMsg$};
\node (H_RMCommittedImpliesNoRMsWorking) at (346.37bp,121.76bp) [draw=black,rectangle,rounded corners] {\huge\emph{RMCommittedImpliesNoRMsWorking}};
\node (H_RMCommittedImpliesNoRMsWorking_RMRcvAbortMsgAction) at (226.37bp,178.41bp) [draw,rectangle,fill=lightgray,proofactionnode] {$RMRcvAbortMsg$};
\node (H_RMCommittedImpliesNoRMsWorking_RMRcvCommitMsgAction) at (346.37bp,178.41bp) [draw,rectangle,fill=lightgray,proofactionnode] {$RMRcvCommitMsg$};
\node (H_RMCommittedImpliesNoAbortMsg) at (155.37bp,235.06bp) [draw=black,rectangle,rounded corners] {\huge\emph{RMCommittedImpliesNoAbortMsg}};
\node (H_RMCommittedImpliesNoAbortMsg_TMAbortAction) at (102.37bp,291.71bp) [draw,rectangle,fill=lightgray,proofactionnode] {$TMAbort$};
\node (H_RMCommittedImpliesNoAbortMsg_RMRcvCommitMsgAction) at (195.37bp,291.71bp) [draw,rectangle,fill=lightgray,proofactionnode] {$RMRcvCommitMsg$};
\node (H_RMCommittedImpliesTMCommitted) at (102.37bp,343.33bp) [draw=black,rectangle,rounded corners] {$L_{0}$};
\node (H_RMCommittedImpliesTMCommitted_RMRcvCommitMsgAction) at (102.37bp,394.95bp) [draw,rectangle,fill=lightgray,proofactionnode] {$RMRcvCommitMsg$};
\node (H_CommitMsgImpliesTMCommitted) at (102.37bp,446.56bp) [draw=black,rectangle,rounded corners] {$L_{1}$};
\node (H_CommitMsgImpliesNoAbortMsg) at (307.37bp,343.33bp) [draw=black,rectangle,rounded corners] {$L_{2}$};
\node (H_CommitSentImpliesRMsNotWorking_RMRcvAbortMsgAction) at (307.37bp,291.71bp) [draw,rectangle,fill=lightgray,proofactionnode] {$RMRcvAbortMsg$};
\node (H_RMAbortedImpliesNoCommitMsg_RMRcvAbortMsgAction) at (413.37bp,291.71bp) [draw,rectangle,fill=lightgray,proofactionnode] {$RMRcvAbortMsg$};
\node (H_CommitMsgImpliesNoAbortMsg_TMAbortAction) at (344.37bp,394.95bp) [draw,rectangle,fill=lightgray,proofactionnode] {$TMAbort$};
\node (H_CommitMsgImpliesNoAbortMsg_TMCommitAction) at (270.37bp,394.95bp) [draw,rectangle,fill=lightgray,proofactionnode] {$TMCommit$};
\node (H_InitNoCommitMsg) at (344.37bp,446.56bp) [draw=black,rectangle,rounded corners] {$L_{3}$};
\node (H_InitNoAbortMsg) at (270.37bp,446.56bp) [draw=black,rectangle,rounded corners] {$L_{4}$};
\node (H_CommitSentImpliesRMsNotWorking) at (346.37bp,235.06bp) [draw=black,rectangle,rounded corners] {$L_{5}$};
\node (H_CommitSentImpliesRMsNotWorking_TMCommitAction) at (506.37bp,291.71bp) [draw,rectangle,fill=lightgray,proofactionnode] {$TMCommit$};
\node (H_RMAbortedAndPreparedImpliesTMAborted) at (540.37bp,343.33bp) [draw=black,rectangle,rounded corners] {$L_{6}$};
\node (H_RMAbortedImpliesNoCommitMsg_TMCommitAction) at (586.37bp,291.71bp) [draw,rectangle,fill=lightgray,proofactionnode] {$TMCommit$};
\node (H_RMAbortedAndPreparedImpliesTMAborted_RMRcvAbortMsgAction) at (620.37bp,394.95bp) [draw,rectangle,fill=lightgray,proofactionnode] {$RMRcvAbortMsg$};
\node (H_RMAbortedAndPreparedImpliesTMAborted_RMChooseToAbortAction) at (517.37bp,394.95bp) [draw,rectangle,fill=lightgray,proofactionnode] {$RMChooseAbort$};
\node (H_TMKnowsPrepareImpliesRMWorking) at (472.37bp,446.56bp) [draw=black,rectangle,rounded corners] {$L_{7}$};
\node (H_TMKnowsPrepareImpliesRMWorking_TMRcvPreparedAction) at (472.37bp,498.18bp) [draw,rectangle,fill=lightgray,proofactionnode] {$TMRcvPrepared$};
\node (H_RMSentPrepareImpliesNotWorking) at (542.37bp,549.8bp) [draw=black,rectangle,rounded corners] {$L_{8}$};
\node (H_RMWorkingImpliesNotPrepared_TMRcvPreparedAction) at (660.37bp,498.18bp) [draw,rectangle,fill=lightgray,proofactionnode] {$TMRcvPrepared$};
\node (H_AbortMsgImpliesTMAborted) at (620.37bp,446.56bp) [draw=black,rectangle,rounded corners] {$L_{9}$};
\node (H_RMAbortedImpliesNoCommitMsg) at (586.37bp,235.06bp) [draw=black,rectangle,rounded corners] {\huge\emph{RMAbortedImpliesNoCommitMsg}};
\node (H_RMAbortedImpliesNoCommitMsg_RMChooseToAbortAction) at (689.37bp,291.71bp) [draw,rectangle,fill=lightgray,proofactionnode] {$RMChooseAbort$};
\node (H_RMWorkingImpliesNoCommitMsg) at (701.37bp,343.33bp) [draw=black,rectangle,rounded corners] {$L_{10}$};
\node (H_RMWorkingImpliesNoCommitMsg_TMCommitAction) at (713.37bp,394.95bp) [draw,rectangle,fill=lightgray,proofactionnode] {$TMCommit$};
\node (H_RMWorkingImpliesNotPrepared) at (685.37bp,446.56bp) [draw=black,rectangle,rounded corners] {$L_{11}$};
\draw [->,proofactionedge] (H_TCConsistent_RMRcvAbortMsgAction) ..controls (238.5bp,49.103bp) and (273.08bp,37.075bp)  .. (H_TCConsistent);
\draw [->,proofactionedge] (H_TCConsistent_RMChooseToAbortAction) ..controls (346.37bp,51.856bp) and (346.37bp,42.953bp)  .. (H_TCConsistent);
\draw [->,proofactionedge] (H_TCConsistent_RMRcvCommitMsgAction) ..controls (532.2bp,96.114bp) and (436.01bp,52.358bp)  .. (H_TCConsistent);
\draw [->,prooflemmaedge] (H_RMCommittedImpliesNoRMsWorking) ..controls (346.37bp,100.8bp) and (346.37bp,91.555bp)  .. (H_TCConsistent_RMChooseToAbortAction);
\draw [->,proofactionedge] (H_RMCommittedImpliesNoRMsWorking_RMRcvAbortMsgAction) ..controls (259.27bp,162.43bp) and (285.95bp,150.27bp)  .. (H_RMCommittedImpliesNoRMsWorking);
\draw [->,proofactionedge] (H_RMCommittedImpliesNoRMsWorking_RMRcvCommitMsgAction) ..controls (346.37bp,164.21bp) and (346.37bp,155.16bp)  .. (H_RMCommittedImpliesNoRMsWorking);
\draw [->,prooflemmaedge] (H_RMCommittedImpliesNoAbortMsg) ..controls (152.67bp,198.51bp) and (151.31bp,148.19bp)  .. (165.37bp,109.05bp) .. controls (168.87bp,99.293bp) and (175.21bp,89.707bp)  .. (H_TCConsistent_RMRcvAbortMsgAction);
\draw [->,prooflemmaedge] (H_RMCommittedImpliesNoAbortMsg) ..controls (182.61bp,213.09bp) and (196.7bp,202.24bp)  .. (H_RMCommittedImpliesNoRMsWorking_RMRcvAbortMsgAction);
\draw [->,proofactionedge] (H_RMCommittedImpliesNoAbortMsg_TMAbortAction) ..controls (115.01bp,277.67bp) and (125.81bp,266.54bp)  .. (H_RMCommittedImpliesNoAbortMsg);
\draw [->,proofactionedge] (H_RMCommittedImpliesNoAbortMsg_RMRcvCommitMsgAction) ..controls (185.3bp,276.95bp) and (177.85bp,266.77bp)  .. (H_RMCommittedImpliesNoAbortMsg);
\draw [->,prooflemmaedge] (H_RMCommittedImpliesTMCommitted) ..controls (102.37bp,328.56bp) and (102.37bp,318.91bp)  .. (H_RMCommittedImpliesNoAbortMsg_TMAbortAction);
\draw [->,proofactionedge] (H_RMCommittedImpliesTMCommitted_RMRcvCommitMsgAction) ..controls (102.37bp,380.4bp) and (102.37bp,371.17bp)  .. (H_RMCommittedImpliesTMCommitted);
\draw [->,prooflemmaedge] (H_CommitMsgImpliesTMCommitted) ..controls (102.37bp,432.01bp) and (102.37bp,422.79bp)  .. (H_RMCommittedImpliesTMCommitted_RMRcvCommitMsgAction);
\draw [->,prooflemmaedge] (H_CommitMsgImpliesNoAbortMsg) ..controls (280.72bp,330.52bp) and (246.44bp,315.34bp)  .. (H_RMCommittedImpliesNoAbortMsg_RMRcvCommitMsgAction);
\draw [->,prooflemmaedge] (H_CommitMsgImpliesNoAbortMsg) ..controls (307.37bp,328.78bp) and (307.37bp,319.55bp)  .. (H_CommitSentImpliesRMsNotWorking_RMRcvAbortMsgAction);
\draw [->,prooflemmaedge] (H_CommitMsgImpliesNoAbortMsg) ..controls (332.92bp,330.37bp) and (364.48bp,315.59bp)  .. (H_RMAbortedImpliesNoCommitMsg_RMRcvAbortMsgAction);
\draw [->,proofactionedge] (H_CommitMsgImpliesNoAbortMsg_TMAbortAction) ..controls (334.84bp,381.17bp) and (326.62bp,370.14bp)  .. (H_CommitMsgImpliesNoAbortMsg);
\draw [->,proofactionedge] (H_CommitMsgImpliesNoAbortMsg_TMCommitAction) ..controls (279.89bp,381.17bp) and (288.12bp,370.14bp)  .. (H_CommitMsgImpliesNoAbortMsg);
\draw [->,prooflemmaedge] (H_InitNoCommitMsg) ..controls (344.37bp,431.8bp) and (344.37bp,422.15bp)  .. (H_CommitMsgImpliesNoAbortMsg_TMAbortAction);
\draw [->,prooflemmaedge] (H_InitNoAbortMsg) ..controls (270.37bp,431.8bp) and (270.37bp,422.15bp)  .. (H_CommitMsgImpliesNoAbortMsg_TMCommitAction);
\draw [->,prooflemmaedge] (H_CommitSentImpliesRMsNotWorking) ..controls (346.37bp,219.12bp) and (346.37bp,207.71bp)  .. (H_RMCommittedImpliesNoRMsWorking_RMRcvCommitMsgAction);
\draw [->,proofactionedge] (H_CommitSentImpliesRMsNotWorking_RMRcvAbortMsgAction) ..controls (318.04bp,275.75bp) and (327.17bp,262.97bp)  .. (H_CommitSentImpliesRMsNotWorking);
\draw [->,proofactionedge] (H_CommitSentImpliesRMsNotWorking_TMCommitAction) ..controls (457.57bp,274.04bp) and (398.44bp,253.84bp)  .. (H_CommitSentImpliesRMsNotWorking);
\draw [->,prooflemmaedge] (H_RMAbortedAndPreparedImpliesTMAborted) ..controls (530.57bp,328.03bp) and (523.22bp,317.3bp)  .. (H_CommitSentImpliesRMsNotWorking_TMCommitAction);
\draw [->,prooflemmaedge] (H_RMAbortedAndPreparedImpliesTMAborted) ..controls (553.83bp,327.8bp) and (564.22bp,316.6bp)  .. (H_RMAbortedImpliesNoCommitMsg_TMCommitAction);
\draw [->,proofactionedge] (H_RMAbortedAndPreparedImpliesTMAborted_RMRcvAbortMsgAction) ..controls (595.71bp,378.65bp) and (574.95bp,365.78bp)  .. (H_RMAbortedAndPreparedImpliesTMAborted);
\draw [->,proofactionedge] (H_RMAbortedAndPreparedImpliesTMAborted_RMChooseToAbortAction) ..controls (523.16bp,381.46bp) and (527.97bp,371.07bp)  .. (H_RMAbortedAndPreparedImpliesTMAborted);
\draw [->,prooflemmaedge] (H_TMKnowsPrepareImpliesRMWorking) ..controls (467.72bp,426.65bp) and (463.79bp,405.02bp)  .. (467.37bp,387.0bp) .. controls (473.09bp,358.21bp) and (487.74bp,327.18bp)  .. (H_CommitSentImpliesRMsNotWorking_TMCommitAction);
\draw [->,prooflemmaedge] (H_TMKnowsPrepareImpliesRMWorking) ..controls (485.54bp,431.04bp) and (495.7bp,419.84bp)  .. (H_RMAbortedAndPreparedImpliesTMAborted_RMChooseToAbortAction);
\draw [->,proofactionedge] (H_TMKnowsPrepareImpliesRMWorking_TMRcvPreparedAction) ..controls (472.37bp,483.63bp) and (472.37bp,474.41bp)  .. (H_TMKnowsPrepareImpliesRMWorking);
\draw [->,prooflemmaedge] (H_RMSentPrepareImpliesNotWorking) ..controls (537.37bp,518.26bp) and (525.64bp,446.51bp)  .. (H_RMAbortedAndPreparedImpliesTMAborted_RMChooseToAbortAction);
\draw [->,prooflemmaedge] (H_RMSentPrepareImpliesNotWorking) ..controls (521.86bp,534.26bp) and (505.14bp,522.41bp)  .. (H_TMKnowsPrepareImpliesRMWorking_TMRcvPreparedAction);
\draw [->,prooflemmaedge] (H_RMSentPrepareImpliesNotWorking) ..controls (569.91bp,537.22bp) and (606.47bp,521.85bp)  .. (H_RMWorkingImpliesNotPrepared_TMRcvPreparedAction);
\draw [->,prooflemmaedge] (H_AbortMsgImpliesTMAborted) ..controls (620.37bp,432.01bp) and (620.37bp,422.79bp)  .. (H_RMAbortedAndPreparedImpliesTMAborted_RMRcvAbortMsgAction);
\draw [->,prooflemmaedge] (H_RMAbortedImpliesNoCommitMsg) ..controls (586.37bp,202.23bp) and (586.37bp,164.18bp)  .. (H_TCConsistent_RMRcvCommitMsgAction);
\draw [->,proofactionedge] (H_RMAbortedImpliesNoCommitMsg_RMRcvAbortMsgAction) ..controls (461.89bp,275.38bp) and (502.8bp,262.46bp)  .. (H_RMAbortedImpliesNoCommitMsg);
\draw [->,proofactionedge] (H_RMAbortedImpliesNoCommitMsg_TMCommitAction) ..controls (586.37bp,278.36bp) and (586.37bp,268.74bp)  .. (H_RMAbortedImpliesNoCommitMsg);
\draw [->,proofactionedge] (H_RMAbortedImpliesNoCommitMsg_RMChooseToAbortAction) ..controls (663.6bp,277.04bp) and (639.8bp,264.41bp)  .. (H_RMAbortedImpliesNoCommitMsg);
\draw [->,prooflemmaedge] (H_RMWorkingImpliesNoCommitMsg) ..controls (698.02bp,328.49bp) and (695.65bp,318.69bp)  .. (H_RMAbortedImpliesNoCommitMsg_RMChooseToAbortAction);
\draw [->,proofactionedge] (H_RMWorkingImpliesNoCommitMsg_TMCommitAction) ..controls (710.38bp,381.6bp) and (707.95bp,371.53bp)  .. (H_RMWorkingImpliesNoCommitMsg);
\draw [->,prooflemmaedge] (H_RMWorkingImpliesNotPrepared) ..controls (693.35bp,431.42bp) and (699.23bp,421.01bp)  .. (H_RMWorkingImpliesNoCommitMsg_TMCommitAction);
\draw [->,proofactionedge] (H_RMWorkingImpliesNotPrepared_TMRcvPreparedAction) ..controls (667.34bp,483.34bp) and (672.28bp,473.53bp)  .. (H_RMWorkingImpliesNotPrepared);
\end{tikzpicture}

%% file: diagrams/AbstractRaft.proof.dot.tex
\begin{tikzpicture}[>=latex,line join=bevel,]

    \Large
\node (H_StateMachineSafety) at (241.14bp,175.96bp) [draw=black,fill=green!50,rectangle] {\huge\emph{StateMachineSafety}};
\node (H_StateMachineSafety_CommitEntryAction) at (214.14bp,232.56bp) [draw,rectangle,fill=lightgray,proofactionnode] {$CommitEntry$};
\node (H_CommittedEntryIsOnQuorum) at (135.14bp,289.15bp) [draw=black,rectangle,rounded corners] {\huge\emph{CommittedEntryIsOnQuorum}};
\node (H_LeaderCompleteness_BecomeLeaderAction) at (480.14bp,232.56bp) [draw,rectangle,fill=lightgray,proofactionnode] {$BecomeLeader$};
\node (H_CommittedEntryIsOnQuorum_RollbackEntriesAction) at (343.14bp,6.979bp) [draw,rectangle,fill=lightgray,proofactionnode] {$RollbackEntries$};
\node (H_LaterLogsHaveEarlierCommitted) at (564.14bp,62.661bp) [draw=black,rectangle,rounded corners] {\huge\emph{LaterLogsHaveEarlierCommitted}};
\node (H_LaterLogsHaveEarlierCommitted_ClientRequestAction) at (471.14bp,119.31bp) [draw,rectangle,fill=lightgray,proofactionnode] {$ClientRequest$};
\node (H_LaterLogsHaveEarlierCommitted_GetEntriesAction) at (602.14bp,232.56bp) [draw,rectangle,fill=lightgray,proofactionnode] {$GetEntries$};
\node (H_LaterLogsHaveEarlierCommitted_CommitEntryAction) at (869.14bp,513.59bp) [draw,rectangle,fill=lightgray,proofactionnode] {$CommitEntry$};
\node (H_LeaderCompleteness) at (443.14bp,175.96bp) [draw=black,rectangle,rounded corners] {\huge\emph{LeaderCompleteness}};
\node (H_LeaderCompleteness_CommitEntryAction) at (389.14bp,232.56bp) [draw,rectangle,fill=lightgray,proofactionnode] {$CommitEntry$};
\node (H_TermsMonotonic) at (436.14bp,289.15bp) [draw=black,rectangle,rounded corners] {\huge\emph{TermsMonotonic}};
\node (H_TermsMonotonic_ClientRequestAction) at (449.14bp,400.41bp) [draw,rectangle,fill=lightgray,proofactionnode] {$ClientRequest$};
\node (H_PrimaryTermGTELogTerm) at (451.14bp,513.59bp) [draw=black,rectangle,rounded corners] {\huge\emph{PrimaryTermGTELogTerm}};
\node (H_PrimaryTermGTELogTerm_BecomeLeaderAction) at (455.14bp,570.13bp) [draw,rectangle,fill=lightgray,proofactionnode] {$BecomeLeader$};
\node (H_LogEntryImpliesSafeAtTerm) at (459.14bp,626.78bp) [draw=black,rectangle,rounded corners] {\huge\emph{LogEntryImpliesSafeAtTerm}};
\node (H_PrimaryHasOwnEntries_BecomeLeaderAction) at (552.14bp,570.13bp) [draw,rectangle,fill=lightgray,proofactionnode] {$BecomeLeader$};
\node (H_LogEntryImpliesSafeAtTerm_ClientRequestAction) at (459.14bp,683.43bp) [draw,rectangle,fill=lightgray,proofactionnode] {$ClientRequest$};
\node (H_QuorumsSafeAtTerms) at (459.14bp,740.08bp) [draw=black,rectangle,rounded corners] {\huge\emph{QuorumsSafeAtTerms}};
\node (H_OnePrimaryPerTerm_BecomeLeaderAction) at (627.14bp,683.43bp) [draw,rectangle,fill=lightgray,proofactionnode] {$BecomeLeader$};
\node (H_UniformLogEntries) at (622.14bp,289.15bp) [draw=black,rectangle,rounded corners] {\huge\emph{UniformLogEntries}};
\node (H_UniformLogEntries_ClientRequestAction) at (706.14bp,457.06bp) [draw,rectangle,fill=lightgray,proofactionnode] {$ClientRequest$};
\node (H_UniformLogEntries_GetEntriesAction) at (617.14bp,344.78bp) [draw,rectangle,fill=lightgray,proofactionnode] {$GetEntries$};
\node (H_LogMatching) at (617.14bp,400.41bp) [draw=black,rectangle,rounded corners] {\huge\emph{LogMatching}};
\node (H_LogMatching_ClientRequestAction) at (617.14bp,457.06bp) [draw,rectangle,fill=lightgray,proofactionnode] {$ClientRequest$};
\node (H_PrimaryHasOwnEntries) at (705.14bp,513.59bp) [draw=black,rectangle,rounded corners] {\huge\emph{PrimaryHasOwnEntries}};
\node (H_PrimaryHasOwnEntries_ClientRequestAction) at (705.14bp,570.13bp) [draw,rectangle,fill=lightgray,proofactionnode] {$ClientRequest$};
\node (H_OnePrimaryPerTerm) at (705.14bp,626.78bp) [draw=black,rectangle,rounded corners] {\huge\emph{OnePrimaryPerTerm}};
\draw [->,proofactionedge] (H_StateMachineSafety_CommitEntryAction) ..controls (220.61bp,218.49bp) and (224.96bp,209.68bp)  .. (H_StateMachineSafety);
\draw [->,prooflemmaedge] (H_CommittedEntryIsOnQuorum) ..controls (165.69bp,267.04bp) and (181.77bp,255.93bp)  .. (H_StateMachineSafety_CommitEntryAction);
\draw [->,prooflemmaedge] (H_CommittedEntryIsOnQuorum) ..controls (282.2bp,264.88bp) and (375.7bp,250.09bp)  .. (H_LeaderCompleteness_BecomeLeaderAction);
\draw [->,proofactionedge] (H_CommittedEntryIsOnQuorum_RollbackEntriesAction) ..controls (238.81bp,16.447bp) and (122.14bp,38.548bp)  .. (122.14bp,118.31bp) .. controls (122.14bp,176.96bp) and (122.14bp,176.96bp)  .. (122.14bp,176.96bp) .. controls (122.14bp,207.3bp) and (126.98bp,241.98bp)  .. (H_CommittedEntryIsOnQuorum);
\draw [->,prooflemmaedge] (H_LaterLogsHaveEarlierCommitted) ..controls (471.75bp,39.218bp) and (416.9bp,25.894bp)  .. (H_CommittedEntryIsOnQuorum_RollbackEntriesAction);
\draw [->,prooflemmaedge] (H_LaterLogsHaveEarlierCommitted) ..controls (567.18bp,100.05bp) and (567.97bp,152.98bp)  .. (545.14bp,188.67bp) .. controls (536.17bp,202.68bp) and (520.8bp,213.14bp)  .. (H_LeaderCompleteness_BecomeLeaderAction);
\draw [->,proofactionedge] (H_LaterLogsHaveEarlierCommitted_ClientRequestAction) ..controls (496.28bp,103.54bp) and (516.2bp,91.836bp)  .. (H_LaterLogsHaveEarlierCommitted);
\draw [->,proofactionedge] (H_LaterLogsHaveEarlierCommitted_GetEntriesAction) ..controls (598.13bp,207.84bp) and (588.82bp,154.76bp)  .. (578.14bp,111.36bp) .. controls (576.17bp,103.35bp) and (573.72bp,94.667bp)  .. (H_LaterLogsHaveEarlierCommitted);
\draw [->,proofactionedge] (H_LaterLogsHaveEarlierCommitted_CommitEntryAction) ..controls (850.65bp,488.61bp) and (820.14bp,444.02bp)  .. (820.14bp,401.41bp) .. controls (820.14bp,401.41bp) and (820.14bp,401.41bp)  .. (820.14bp,174.96bp) .. controls (820.14bp,121.8bp) and (773.14bp,93.766bp)  .. (H_LaterLogsHaveEarlierCommitted);
\draw [->,prooflemmaedge] (H_LeaderCompleteness) ..controls (453.37bp,155.0bp) and (458.11bp,145.74bp)  .. (H_LaterLogsHaveEarlierCommitted_ClientRequestAction);
\draw [->,proofactionedge] (H_LeaderCompleteness_BecomeLeaderAction) ..controls (471.5bp,218.81bp) and (464.38bp,208.3bp)  .. (H_LeaderCompleteness);
\draw [->,proofactionedge] (H_LeaderCompleteness_CommitEntryAction) ..controls (403.01bp,217.54bp) and (413.65bp,206.78bp)  .. (H_LeaderCompleteness);
\draw [->,prooflemmaedge] (H_TermsMonotonic) ..controls (498.09bp,267.78bp) and (542.49bp,253.18bp)  .. (H_LaterLogsHaveEarlierCommitted_GetEntriesAction);
\draw [->,prooflemmaedge] (H_TermsMonotonic) ..controls (451.23bp,269.43bp) and (460.48bp,257.95bp)  .. (H_LeaderCompleteness_BecomeLeaderAction);
\draw [->,prooflemmaedge] (H_TermsMonotonic) ..controls (420.19bp,269.62bp) and (410.6bp,258.48bp)  .. (H_LeaderCompleteness_CommitEntryAction);
\draw [->,proofactionedge] (H_TermsMonotonic_ClientRequestAction) ..controls (446.38bp,376.16bp) and (441.68bp,336.66bp)  .. (H_TermsMonotonic);
\draw [->,prooflemmaedge] (H_PrimaryTermGTELogTerm) ..controls (450.57bp,480.89bp) and (449.89bp,443.16bp)  .. (H_TermsMonotonic_ClientRequestAction);
\draw [->,proofactionedge] (H_PrimaryTermGTELogTerm_BecomeLeaderAction) ..controls (454.24bp,556.8bp) and (453.53bp,547.21bp)  .. (H_PrimaryTermGTELogTerm);
\draw [->,prooflemmaedge] (H_LogEntryImpliesSafeAtTerm) ..controls (535.57bp,599.56bp) and (601.98bp,577.73bp)  .. (660.14bp,562.18bp) .. controls (732.23bp,542.91bp) and (751.55bp,543.49bp)  .. (824.14bp,526.18bp) .. controls (826.12bp,525.71bp) and (828.14bp,525.22bp)  .. (H_LaterLogsHaveEarlierCommitted_CommitEntryAction);
\draw [->,prooflemmaedge] (H_LogEntryImpliesSafeAtTerm) ..controls (457.68bp,605.83bp) and (457.0bp,596.58bp)  .. (H_PrimaryTermGTELogTerm_BecomeLeaderAction);
\draw [->,prooflemmaedge] (H_LogEntryImpliesSafeAtTerm) ..controls (496.08bp,604.07bp) and (516.61bp,592.01bp)  .. (H_PrimaryHasOwnEntries_BecomeLeaderAction);
\draw [->,proofactionedge] (H_LogEntryImpliesSafeAtTerm_ClientRequestAction) ..controls (459.14bp,669.23bp) and (459.14bp,660.18bp)  .. (H_LogEntryImpliesSafeAtTerm);
\draw [->,prooflemmaedge] (H_QuorumsSafeAtTerms) ..controls (359.86bp,711.47bp) and (298.14bp,681.64bp)  .. (298.14bp,627.78bp) .. controls (298.14bp,627.78bp) and (298.14bp,627.78bp)  .. (298.14bp,343.78bp) .. controls (298.14bp,300.65bp) and (340.08bp,265.28bp)  .. (H_LeaderCompleteness_CommitEntryAction);
\draw [->,prooflemmaedge] (H_QuorumsSafeAtTerms) ..controls (459.14bp,719.54bp) and (459.14bp,710.92bp)  .. (H_LogEntryImpliesSafeAtTerm_ClientRequestAction);
\draw [->,prooflemmaedge] (H_QuorumsSafeAtTerms) ..controls (528.27bp,716.6bp) and (569.42bp,703.21bp)  .. (H_OnePrimaryPerTerm_BecomeLeaderAction);
\draw [->,prooflemmaedge] (H_UniformLogEntries) ..controls (614.78bp,268.04bp) and (611.29bp,258.53bp)  .. (H_LaterLogsHaveEarlierCommitted_GetEntriesAction);
\draw [->,prooflemmaedge] (H_UniformLogEntries) ..controls (564.42bp,265.96bp) and (530.85bp,253.06bp)  .. (H_LeaderCompleteness_BecomeLeaderAction);
\draw [->,proofactionedge] (H_UniformLogEntries_ClientRequestAction) ..controls (701.97bp,435.84bp) and (695.63bp,408.92bp)  .. (686.14bp,387.7bp) .. controls (675.49bp,363.89bp) and (669.46bp,359.67bp)  .. (655.14bp,337.86bp) .. controls (649.49bp,329.24bp) and (643.08bp,319.88bp)  .. (H_UniformLogEntries);
\draw [->,proofactionedge] (H_UniformLogEntries_GetEntriesAction) ..controls (618.26bp,331.8bp) and (619.11bp,322.68bp)  .. (H_UniformLogEntries);
\draw [->,prooflemmaedge] (H_LogMatching) ..controls (617.14bp,379.99bp) and (617.14bp,371.17bp)  .. (H_UniformLogEntries_GetEntriesAction);
\draw [->,proofactionedge] (H_LogMatching_ClientRequestAction) ..controls (617.14bp,442.86bp) and (617.14bp,433.81bp)  .. (H_LogMatching);
\draw [->,prooflemmaedge] (H_PrimaryHasOwnEntries) ..controls (705.5bp,493.21bp) and (705.66bp,484.3bp)  .. (H_UniformLogEntries_ClientRequestAction);
\draw [->,prooflemmaedge] (H_PrimaryHasOwnEntries) ..controls (671.22bp,491.57bp) and (652.61bp,480.04bp)  .. (H_LogMatching_ClientRequestAction);
\draw [->,proofactionedge] (H_PrimaryHasOwnEntries_ClientRequestAction) ..controls (705.14bp,555.96bp) and (705.14bp,546.93bp)  .. (H_PrimaryHasOwnEntries);
\draw [->,proofactionedge] (H_PrimaryHasOwnEntries_BecomeLeaderAction) ..controls (592.08bp,554.89bp) and (629.4bp,541.59bp)  .. (H_PrimaryHasOwnEntries);
\draw [->,prooflemmaedge] (H_OnePrimaryPerTerm) ..controls (705.14bp,606.35bp) and (705.14bp,597.43bp)  .. (H_PrimaryHasOwnEntries_ClientRequestAction);
\draw [->,proofactionedge] (H_OnePrimaryPerTerm_BecomeLeaderAction) ..controls (646.33bp,668.99bp) and (663.59bp,656.89bp)  .. (H_OnePrimaryPerTerm);
\end{tikzpicture}